# A Unified Energy-Density Framework for Moiré Interface Energetics and Mechanics

Zhangke Lei[1], Weidong Yan[2], Zhicheng Feng[3], Wengen Ouyang[1,4*]

[1]*Department of Engineering Mechanics, School of Civil Engineering, Wuhan University, Wuhan, Hubei 430072, China*

[2]*School of Mechanical and Electrical Engineering, Wuhan Institute of Technology, Wuhan, China*

[3]*Department of Engineering Mechanics, Applied Mechanics Lab, Tsinghua University, Beijing, 100084, P.R.China*

[4]*State Key Laboratory of Water Resources Engineering and Management, Wuhan University, Wuhan, Hubei 430072, China*

[*]Corresponding author: w.g.ouyang@whu.edu.cn

**Abstract**

The mechanical response of moiré interfaces depends on how atomic registry combines with contact size and boundary geometry. An analytical description must retain this discrete lattice information while connecting microscopic interactions to the energetics of finite contacts. Here, we develop a unified energy-density framework for rigid van der Waals interfaces by expressing registry-dependent pair potentials and atomic densities in reciprocal space. The resulting formulation accommodates different lattice bases and atomic species, linking single-atom and molecular adsorption to infinite bilayers and finite crystalline contacts. Explicit expressions for circular and regular polygonal flakes separate the area contribution from boundary-dependent terms and connect microscopic corrugation coefficients to energy barriers, rotational torques, and interlayer shear stiffnesses. The boundary analysis recovers the quarter-power area scaling of the barrier envelope for circular flakes and distinguishes bounded oscillations from square-root area scaling for polygonal contacts according to edge orientation. A dimensionless criterion characterizes the polygon-to-circle crossover in terms of moiré wavelength, flake size, and side number. Direct-summation benchmarks support the finite-domain expressions, while calculations for two-dimensional ice on graphene demonstrate the corresponding boundary-scaling trends in a molecular lattice. Applications to interlayer shear and graphite compression connect the same microscopic interaction description to tangential and normal mechanical responses. The framework establishes an analytical connection between microscopic registry, moiré periodicity, and finite-domain geometry, providing a basis for understanding and tailoring the energetics and mechanics of van der Waals contacts.

## 1 Introduction

Twisted interfaces between crystalline solids give rise to moiré superlattices [1,2] that profoundly alter interfacial energetics and mechanical response [3-5]. In layered van der Waals (vdW) materials, rotational misalignment between adjacent lattices generates long-wavelength modulation of local stacking configurations (moiré patterns), producing spatially varying interaction energies even in the absence of chemical heterogeneity. These moiré superlattices are now recognized as a powerful design principle, governing phenomena ranging from structural superlubricity [6-20], twistronics [21-23], anisotropic thermal transport [24-27], to moiré-tunable mechanical properties [5,28]. From a mechanics perspective, these systems present a unique paradigm for understanding how geometric misfit and finite-domain boundaries govern energy landscapes across atomic to continuum scales.

Early studies of twisted interfaces primarily relied on atomistic simulations or experiments to characterize frictional forces and energy barriers, revealing unconventional scaling laws with contact area and twist angle [29-31]. The observed scaling includes transitions between sublinear and linear regimes governed by commensurability and contact geometry [14,29]. While invaluable, such numerical approaches often obscure the underlying physical mechanisms and offer limited predictive power beyond simulated specifics. Theoretical works have further highlighted that geometry and orientation can qualitatively alter this scaling behavior [15,32,33]. For instance, circular and polygonal flakes exhibit distinct power laws, suggesting competition between edge effects and bulk moiré modulation. The edge effects itself have been analyzed, for polygonal flakes, to determine the conditions under which sliding becomes frictionless [34,35]. A common description of circular and regular-polygonal contacts would make the boundary contribution an explicit geometric variable and connect their distinct scaling regimes. This motivates a unified analytical treatment linking microscopic interactions, moiré periodicity and boundary geometry, together with a quantitative criterion for the polygon-to-circle crossover.

Continuum approaches connect atomistic interactions to effective energy densities [36-38]. For finite moiré contacts, this connection must preserve both the registry dependence of the local energy and the geometry of the integration domain. Moreover, they often fail to incorporate the essential moiré-modulated energy landscape that encodes stacking variations. Recent efforts to include moiré effects [39] motivate an analytical formulation in which lattice registry and finite-domain geometry enter explicitly.

In this work, we develop a unified energy-density framework that connects microscopic registry-dependent pair interactions to the energetics and mechanics of finite moiré interfaces. By expressing interatomic interactions and atomic densities in reciprocal space, the formulation links microscopic Fourier coefficients to finite-domain integrals for arbitrary contact geometries on an infinite substrate, cleanly separating the bulk area contribution from boundary-governed terms. This common analytical representation recovers the classical Bessel form for circular flakes and yields explicit boundary expressions for regular polygonal contacts, allowing their orientation-dependent scaling behaviors and the polygon-to-circle crossover to be analyzed within a single framework. Energy derivatives with respect to translation, rotation, and interlayer spacing then connect these geometric terms directly to conservative lateral forces, rotational torques, and interfacial shear and normal stiffnesses.

Different lattice bases and pair potentials enter the formulation modularly through their corresponding Fourier coefficients, providing a general route from single-atom and molecular adsorption to infinite twisted bilayers and finite crystalline contacts. Using graphitic interfaces modeled by the Kolmogorov–Crespi potential as a foundational benchmark, we demonstrate how moiré periodicity and periphery geometry jointly govern the interfacial energy landscape across circular and polygonal flakes. Moreover, by representing molecular orientation via local normal vectors, we capture the binding and sliding energetics of a single water molecule on graphene [40] and, without additional fitting, predict the finite-domain energetics of two-dimensional ice on graphene [41]. The calculated barrier amplitudes for 2D ice across different boundary shapes follow the analytical scaling laws of our theory, providing a strict bottom-up consistency test on a molecular lattice entirely distinct from graphene. While layer relaxation, elastic reconstruction, dislocation nucleation, thermal fluctuations, and dissipative friction lie beyond the strict rigid-lattice limit, they ultimately build upon the underlying rigid-interface interaction landscape. By establishing an explicit analytical bridge between microscopic registry, moiré periodicity, and contact boundary geometry, our framework provides both a quantitative design tool for rigid crystalline contacts and a foundational baseline for controlling the broader interfacial mechanics of van der Waals heterostructures.

# 2 Theoretical framework

## 2.1 Physical interaction of a single atom/molecule with 2D substrates

The framework expresses substrate periodicity through reciprocal lattice vectors and microscopic interactions through Fourier coefficients. This representation connects single-atom and molecular adsorption to the energy of infinite bilayers and finite crystalline flakes.

### *2.1.1 General Framework for Periodic 2D Systems*

Consider an infinite, rigid two-dimensional crystal substrate defined by lattice vectors $\boldsymbol{a}_1, \boldsymbol{a}_2$, any lattice-dependent physical field $u(\boldsymbol{r})$ satisfies (see Fig. 1),

$$u(\boldsymbol{r}+\boldsymbol{L}) = u(\boldsymbol{r}), \qquad \boldsymbol{L} = l_1\boldsymbol{a}_1 + l_2\boldsymbol{a}_2, l_1, l_2 \in \mathbb{Z} \tag{1}$$

where $\boldsymbol{r} = (\boldsymbol{\rho}, z)$ denotes spatial coordinates ($\boldsymbol{\rho}$: in-plane, $z$: out-of-plane). $\boldsymbol{L}$ is lattice vector, a linear combination of $\boldsymbol{a}_1, \boldsymbol{a}_2$. This periodic field admits a Fourier series expansion [42]:

$$u(\boldsymbol{\rho}, z) = \sum_{\boldsymbol{n}} C_{\boldsymbol{n}}(z) \exp(i\boldsymbol{G}\cdot\boldsymbol{\rho}). \tag{2}$$

Here, Fourier coefficient $C_{\boldsymbol{n}}$ depends on the explicit form of the physical field $u(\boldsymbol{r})$, $\boldsymbol{G} = n_1\boldsymbol{b}_1 + n_2\boldsymbol{b}_2$ is a linear combination of reciprocal lattice vectors $\boldsymbol{b}_1, \boldsymbol{b}_2$ satisfying:

$$\boldsymbol{a}_\mu \cdot \boldsymbol{b}_\nu = 2\pi\delta_{\mu\nu}, \qquad \forall \mu, \nu \in \{1,2\}. \tag{3}$$

The reciprocal lattice vector in Eq. (2) involves two independent integer indices $n_1$ and $n_2$, for simplicity, we combine these two summation variables into a single vector notation $\boldsymbol{n} = (n_1, n_2)$. To compute the potential energy of a single atom at position $(\boldsymbol{\rho}, z)$ interacting with an infinite 2D substrate, we employ pairwise summation over all substrate atoms [42]:

$$u(\boldsymbol{\rho}, z) = \sum_{\boldsymbol{l}} \sum_{k} u_{\text{atom}}(\boldsymbol{\rho} - (\boldsymbol{L} + \boldsymbol{B}_k), z)\,, \tag{4}$$

where $u_{\text{atom}}$ is the interatomic potential, $\boldsymbol{B}_k$ denotes the coordinates of the $k^{\text{th}}$ atom in a unit cell lattice, and $\boldsymbol{L}$ sums over lattice sites. Fig. 1 plots the physical quantities discussed above.

Reciprocal space provides a compact representation of this infinite real-space sum. Under the assumptions of an infinite and rigid substrate, the $u(\boldsymbol{\rho}, z)$ satisfies discrete lattice translational symmetry (Eq. (1)). Substituting Eq. (4) into the Fourier expansion (Eq. (2)) gives the coefficients by Steele's method [42] (see Sec. A.1 of the Supporting Information (SI)):

$$C_{\boldsymbol{n}}(z) = \frac{V_{\boldsymbol{n}}(z)}{a_s} \left[ \sum_{k} \exp(-i\boldsymbol{G} \cdot \boldsymbol{B}_k) \right], \tag{5}$$

where $a_s$ is the area of one Bravais lattice and $V_{\boldsymbol{n}}(z)$ reads as:

$$V_{\boldsymbol{n}}(z) = \int_{S_\infty} u_{\text{atom}}(\boldsymbol{X}, z) \exp(-i\boldsymbol{G} \cdot \boldsymbol{X})\, \mathrm{d}\boldsymbol{X} \tag{6}$$

where the variable substitution $\boldsymbol{X} = \boldsymbol{\rho} - \boldsymbol{L} - \boldsymbol{B}_k$ was used and the integral domain $S_\infty$ represents the entire $xy$ plane. Substituting Eq. (5) to Eq. (2), we have

$$u(\boldsymbol{\rho}, z) = \frac{1}{a_s} \sum_{\boldsymbol{n}} \left[ \sum_{k} e^{-i\boldsymbol{G} \cdot \boldsymbol{B}_k} \right] V_{\boldsymbol{n}}(z) e^{i\boldsymbol{G} \cdot \boldsymbol{\rho}}\,, \tag{7}$$

The resulting expression separates the substrate geometry from the pair interaction, making the single-atom potential directly evaluable in reciprocal space.

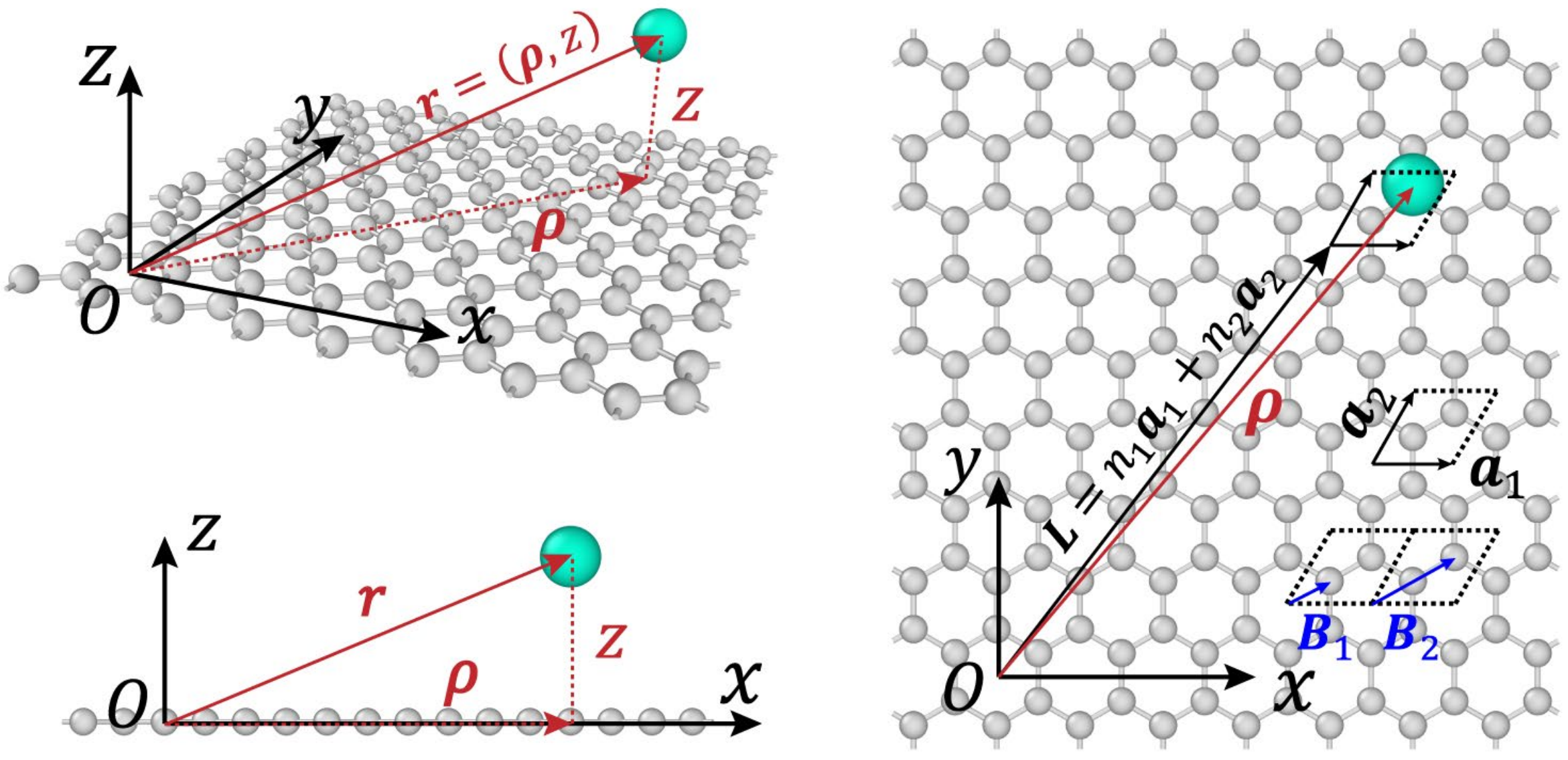

Fig. 1. Schematic of the single atom–substrate physical model. The lattice basis vectors $\boldsymbol{a}_1, \boldsymbol{a}_2$ and the relative position vector of the atom in the Bravais lattice $\boldsymbol{B}$ are indicated.

### *2.1.2 Single-Atom Adsorption on Graphene*

We now apply the above formalism to describe the interaction of a single carbon atom and an infinite graphene substrate using the registry-dependent Kolmogorov-Crespi (KC) potential [43] with the atomic normal vector $\boldsymbol{n}_{\text{atom}}$ and $\boldsymbol{n}_{\text{sub}}$, which captures the anisotropy of carbon-carbon interactions. Here for simplicity, we set the normal vector $\boldsymbol{n}_{\text{atom}} = \boldsymbol{n}_{\text{sub}} = (0,0,1)$ to get atomic KC-z potential:

$$u_{\text{atom}}(\rho, z) = \left\{c + 2\left[c_0 + c_2\left(\frac{\rho}{\delta}\right)^2 + c_4\left(\frac{\rho}{\delta}\right)^4\right]e^{-\left(\frac{\rho}{\delta}\right)^2}\right\}e^{-\lambda\left(\sqrt{\rho^2+z^2}-z_0\right)} - \frac{Az_0^6}{(\rho^2+z^2)^3}, \qquad (8)$$

The analytical evaluation of Eq. (6) for KC-z potential yields (see SI A.1 for derivation details):

$$V_{\boldsymbol{n}}(z) = 2\pi e^{\lambda(z_0-z)}\left\{\frac{c}{2\mu_0|\boldsymbol{G}|^2}e^{-\frac{1}{4\mu_0}} + 2e^{-\frac{1}{4\mu}}\left[\frac{c_0}{2\mu|\boldsymbol{G}|^2} - \frac{c_2(1-4\mu)}{8\mu^3\delta^2|\boldsymbol{G}|^4} + \frac{c_4(32\mu^2-16\mu+1)}{32\mu^5\delta^4|\boldsymbol{G}|^6}\right]\right\}$$
$$-\frac{\pi A z_0^6}{4z^2}|\boldsymbol{G}|^2 K_2(|\boldsymbol{G}|z). \qquad (9)$$

Here integers $\boldsymbol{n} = (n_1, n_2)$ specify the reciprocal lattice vector as $\boldsymbol{G} = n_1\boldsymbol{b}_1 + n_2\boldsymbol{b}_2$. The two dimensionless parameters $\mu_0, \mu$ are defined as

$$\mu_0 = \frac{\lambda}{2z}\frac{1}{|\boldsymbol{G}|^2}, \mu = \left(\frac{\lambda}{2z} + \frac{1}{\delta^2}\right)\frac{1}{|\boldsymbol{G}|^2}, \qquad (10)$$

and function $K_2$ is the modified Bessel function of the second kind.

The graphene lattice is denoted by lattice vectors $\boldsymbol{a}_1 = (a, 0)$, $\boldsymbol{a}_2 = a\left(1/2, \sqrt{3}/2\right)$ with lattice constant $a = \sqrt{3}a_{\text{CC}} \approx 2.46$ Å, which is from the reference density functional theory (DFT) data for bilayer graphene [44]. The corresponding reciprocal lattice vectors, obtained via Eq. (3), are $\boldsymbol{b}_1 = \frac{4\pi}{\sqrt{3}a}\left(\sqrt{3}/2, -1/2\right)$, $\boldsymbol{b}_2 = \frac{4\pi}{\sqrt{3}a}(0,1)$. By truncating the Fourier series in Eq. (7) to the first-order ($|\boldsymbol{G}| = |\boldsymbol{b}_1| = |\boldsymbol{b}_2|$), we obtain a closed-form expression for the interaction potential of a single atom on a graphene substrate:

$$u(\boldsymbol{\rho}, z) = \frac{2V_0(z)}{a_s} - \frac{2V_1(z)}{a_s}\left[\cos(\boldsymbol{b}_1\cdot\boldsymbol{\rho}) + \cos(\boldsymbol{b}_2\cdot\boldsymbol{\rho}) + \cos\left((\boldsymbol{b}_1+\boldsymbol{b}_2)\cdot\boldsymbol{\rho}\right)\right]. \qquad (11)$$

Here, $V_0(z) = V_{\boldsymbol{n}}(z; |\boldsymbol{G}| \to 0), V_1(z) = V_{\boldsymbol{n}}(z; |\boldsymbol{G}| = |\boldsymbol{b}_1| = |\boldsymbol{b}_2|)$ and $a_s = a^2\sin(\pi/3) = \sqrt{3}a^2/2$ is the area of one Bravais lattice. Using the fitted KC parameters and fixing the interlayer spacing at the more commonly used experimental value of 3.35 Å [45,46], we obtain $V_0(3.35) = -94.732\ \text{meV}\cdot\text{Å}^2$ and $V_1(3.35) = 17.886\ \text{meV}\cdot\text{Å}^2$. Substituting the specific form of the reciprocal lattice vector yields the corresponding expression in terms of the $x$ and $y$ coordinates

$$u(\boldsymbol{\rho}, z) = \frac{2V_0(z)}{a_s} - \frac{2V_1(z)}{a_s}\left[\cos\left(\frac{4\pi}{\sqrt{3}a}y\right) + 2\cos\left(\frac{2\pi}{a}x\right)\cos\left(\frac{2\pi}{\sqrt{3}a}y\right)\right], \qquad (12)$$

a form commonly used in previous works[14,15,33,47], in which the formula usually contains a constant term $U_0$ that needs to be fitted to calibrate the energy amplitude at a fixed height. Here the amplitude and its height dependence follow directly from the specified KC potential, without an additional amplitude fit.

As shown in Fig. 2 , we compare sliding potential energy surface computed at 3.4 Å – the interlayer spacing used in the DFT calculations – with the binding energy of bilayer graphene obtained from DFT, the fitting results using the KC-z potential, and the calculations based on our single-atom formula, Eq. (11). The comparison tests how the analytical expression reproduces the binding curve and sliding corrugation represented by the fitted KC-z potential. Because the DFT results report the interaction energy averaged per atom, whereas our theoretical framework calculates the interaction energy of a single atom with the substrate, we evaluate the single-atom formula for both inequivalent atoms in the primitive unit cell. By substituting the relevant data and averaging these two contributions, we obtain the explicit expression for the per-atom interaction energy:

$$w_{\mathrm{AB}} = \frac{1}{4}\left(u(\boldsymbol{\rho}, z) + u\left[\boldsymbol{\rho} + \frac{2}{3}(\boldsymbol{a}_1 + \boldsymbol{a}_2), z\right]\right)$$
$$= \frac{V_0(z)}{a_{\mathrm{s}}} + \frac{V_1(z)}{2a_{\mathrm{s}}}\left[\cos\left(\boldsymbol{b}_1 \cdot \boldsymbol{\rho} + \frac{2\pi}{3}\right) + \cos\left(\boldsymbol{b}_2 \cdot \boldsymbol{\rho} + \frac{2\pi}{3}\right) + \cos\left((\boldsymbol{b}_1 + \boldsymbol{b}_2) \cdot \boldsymbol{\rho} + \frac{4\pi}{3}\right)\right]. \quad (13)$$

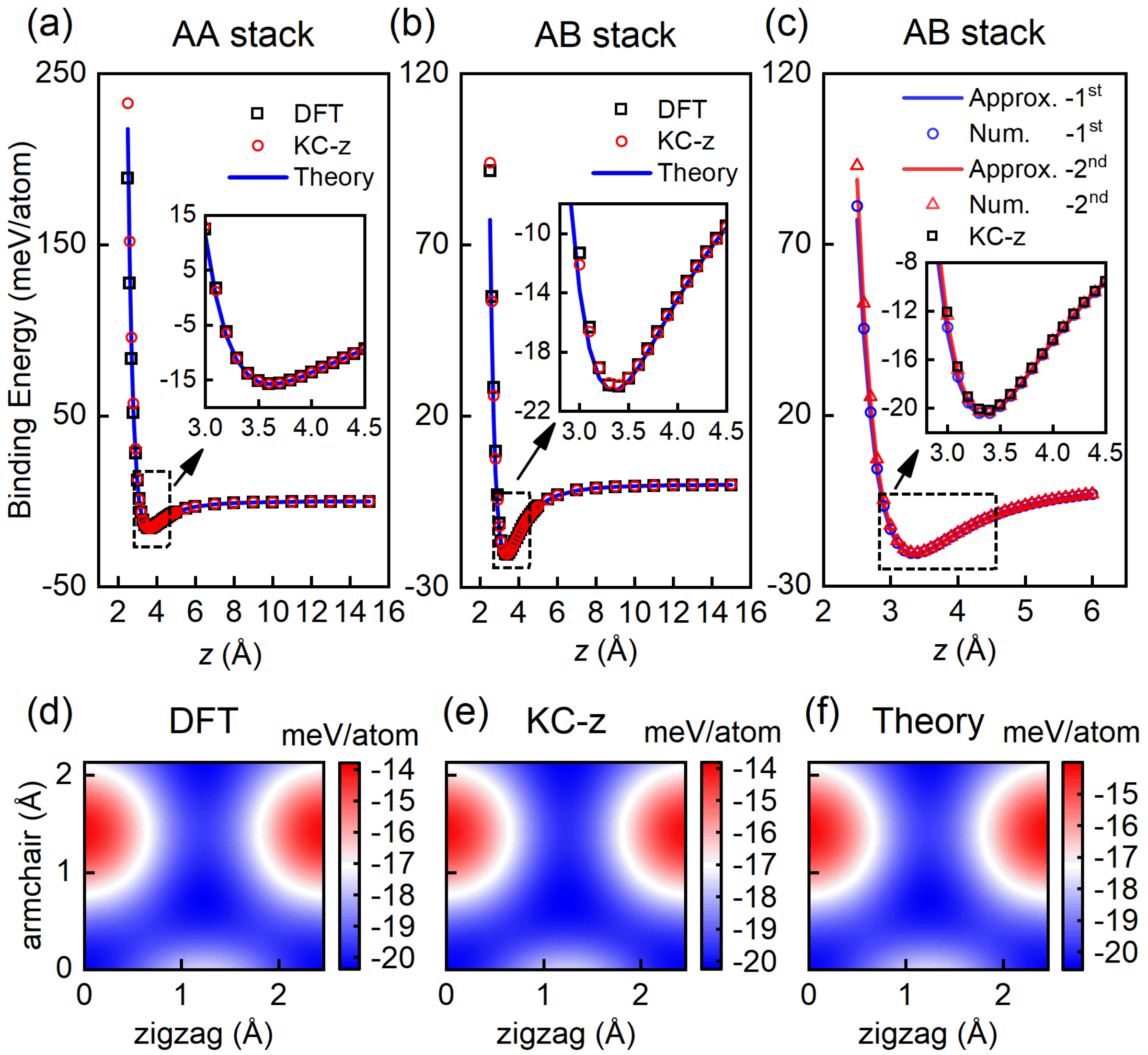

Fig. 2. (a-b) Binding energy of bilayer graphene from DFT, KC-z fitting, and our theoretical calculations. Inset: magnified view of the region indicated by the black square, showing the comparison near the equilibrium height; (c) Validation of $1^{st}$-order formula (blue curves) and $2^{nd}$-order formula (red curves). Comparison between the results from numerical integration (scattered points) of Eq. (6) and the analytical solution (solid lines) Eq. (9) obtained with the approximated KC potential, evaluated at different heights. The $1^{th}$-order and $2^{nd}$-order results are both very close to the fitting results of KC-z near the equilibrium position of the binding energy; (d-f) Sliding energy surface of bilayer graphene at a height of 3.4 Å.

This per-atom expression is used for the parameter fitting in Fig. 2. The same figure also assesses the approximations used in the analytical energy. A direct comparison between the approximate analytical expression (Eq. (9)) and the exact numerical integration of Eq. (6) demonstrates the high accuracy of the approximation. Additionally, we assess the convergence of the Fourier expansion by comparing the second-order truncation of Eq.(7) (detailed in SI A.1) with the first-order approximation given by Eq.(11). As illustrated in Fig. 2c, the discrepancy between the first- and second-order models is negligible. The close agreement supports the first-order truncation for the conditions examined here.

### *2.1.3 Extensions to Other Lattices*

The derivation above relies only on the translational symmetry of the substrate; the potential energy of the atom on the substrate enters exclusively through (i) the reciprocal vectors $\boldsymbol{G}$ of its Bravais lattice, (ii) the in-plane positions $\boldsymbol{B}_k$ of the atoms in the primitive cell, and (iii) the pair potential $u_{\mathrm{atom}}$. Allowing different substrate atoms to have distinct pair interactions $u_{\mathrm{atom}}^{(k)}$ gives the generalized form

$$u(\boldsymbol{\rho}, z) = \frac{1}{a_s}\sum_{\boldsymbol{n}}\sum_{k}\left[e^{-i\boldsymbol{G}\cdot\boldsymbol{B}_k} V_{\boldsymbol{n}}^{(k)}(z - z_k)\right] e^{i\boldsymbol{G}\cdot\boldsymbol{\rho}}, \qquad V_{\boldsymbol{n}}^{(k)}(z) = \int_{S_\infty} u_{\mathrm{atom}}^{(k)}(\boldsymbol{X}, z)\exp(-i\boldsymbol{G}\cdot\boldsymbol{X})\,\mathrm{d}\boldsymbol{X} \quad (14)$$

which reduces to Eq. (7) when all atoms share one species $u_{\mathrm{atom}}$ and one plane. The following lattice examples show how the basis and atomic species enter the same representation:

(i) Hexagonal lattices. For the honeycomb lattice of graphene treated above, the structure factor $\sum_k e^{-i\boldsymbol{G}\cdot\boldsymbol{B}_k} = 2$ for the constant term (two atoms in the primitive cell) and $\sum_k e^{-i\boldsymbol{G}\cdot\boldsymbol{B}_k} = -1$ for the $1^{\mathrm{st}}$-order. For one atom per primitive cell, as in a close-packed metal monolayer (Au (111)), $\sum_k e^{-i\boldsymbol{G}\cdot\boldsymbol{B}_k} = 1$ for the constant term and $\sum_k e^{-i\boldsymbol{G}\cdot\boldsymbol{B}_k} = 1$ for the $1^{\mathrm{st}}$-order, which gives

$$u(\boldsymbol{\rho}, z) = \frac{V_0(z)}{a_s} + \frac{2V_1(z)}{a_s}\left[\cos(\boldsymbol{b}_1 \cdot \boldsymbol{\rho}) + \cos(\boldsymbol{b}_2 \cdot \boldsymbol{\rho}) + \cos\left((\boldsymbol{b}_1 + \boldsymbol{b}_2) \cdot \boldsymbol{\rho}\right)\right] \quad (15)$$

with the single atom placed at the origin, $B_1 = (0,0)$.

(ii) Square Bravais lattices. For the square net of an fcc (100) face, one atom per primitive cell, $\boldsymbol{B}_1 = (0,0)$ and the $1^{\mathrm{st}}$-order reciprocal lattice vectors are $\{\pm\boldsymbol{b}_1, \pm\boldsymbol{b}_2\}$ with $|\boldsymbol{b}_1| = |\boldsymbol{b}_2| = 2\pi/a$, the Eq. (7) gives

$$u(x, y, z) = \frac{V_0(z)}{a_s} + \frac{2V_1(z)}{a_s}\left[\cos\left(\frac{2\pi}{a}x\right) + \cos\left(\frac{2\pi}{a}y\right)\right]. \quad (16)$$

(iii) Multi-species hexagonal lattices. For hexagonal boron nitride (h-BN) the two sublattices are occupied by B and N and no longer equivalent, so the potential is the sum of two hexagonal lattices displaced by $\boldsymbol{B}_{\mathrm{N}}=\frac{1}{3}(\boldsymbol{a}_1,\boldsymbol{a}_2)$ and $\boldsymbol{B}_{\mathrm{B}}=\frac{2}{3}(\boldsymbol{a}_1,\boldsymbol{a}_2)$, and Eq. (7) gives

$$u(\boldsymbol{\rho},z)=\frac{V_0^{\mathrm{N}}(z)+V_0^{\mathrm{B}}(z)}{a_s}+\frac{2V_1^{\mathrm{N}}(z)}{a_s}\left[\cos\left(\boldsymbol{b}_1\cdot\boldsymbol{\rho}-\frac{2\pi}{3}\right)+\cos\left(\boldsymbol{b}_2\cdot\boldsymbol{\rho}-\frac{2\pi}{3}\right)+\cos\left((\boldsymbol{b}_1+\boldsymbol{b}_2)\cdot\boldsymbol{\rho}+\frac{2\pi}{3}\right)\right]+\frac{2V_1^{\mathrm{B}}(z)}{a_s}\left[\cos\left(\boldsymbol{b}_1\cdot\boldsymbol{\rho}+\frac{2\pi}{3}\right)+\cos\left(\boldsymbol{b}_2\cdot\boldsymbol{\rho}+\frac{2\pi}{3}\right)+\cos\left((\boldsymbol{b}_1+\boldsymbol{b}_2)\cdot\boldsymbol{\rho}-\frac{2\pi}{3}\right)\right] \tag{17}$$

here we use $V_{\boldsymbol{n}}^{\mathrm{B}}(z)$ and $V_{\boldsymbol{n}}^{\mathrm{N}}(z)$ to denote the atomic interaction potentials corresponding to the B and N atoms, respectively.

(iv) Lattices with a finite out-of-plane extent. For a transition-metal dichalcogenide such as $MoS_2$ the primitive cell contains a Mo plane flanked by two S planes at heights $h_{\mathrm{Mo}}=0$ and $h_{\mathrm{S}}=\pm h$ (with z measured from the Mo plane), and the three sites carry different pair potentials. Each plane contributes its own Eq. (7), evaluated at the displaced height $z-h_k$ and at its own in-plane position $\boldsymbol{B}_k$:

$$u(\boldsymbol{\rho},z)=\frac{1}{a_s}\sum_k\left[V_0^{(k)}(z-h_k)+2V_1^{(k)}(z-h_k)\sum_i^3\cos\big(\boldsymbol{b}_i\cdot(\boldsymbol{\rho}-\boldsymbol{B}_k)\big)\right], \tag{18}$$

The index $k$ refers to the Mo atom and the two S atoms.

(v) Quasiperiodic substrates. When Eq. (1) fails, Eq. (2) is inapplicable; since Eq. (4) is linear in the substrate atoms, a quasiperiodic surface can nevertheless be assembled by superposing rotated copies of a periodic one:

$$u_{\mathrm{QC}}(\boldsymbol{\rho},z)=\frac{4V_0(z)}{a_{\mathrm{s}}}-\frac{2V_1(z)}{a_{\mathrm{s}}}\big[\cos(\boldsymbol{b}_1\cdot\boldsymbol{\rho})+\cos(\boldsymbol{b}_2\cdot\boldsymbol{\rho})+\cos\big((\boldsymbol{b}_1+\boldsymbol{b}_2)\cdot\boldsymbol{\rho}\big)\big]-\frac{2V_1(z)}{a_{\mathrm{s}}}\big[\cos(\boldsymbol{b}_1'\cdot\boldsymbol{\rho})+\cos(\boldsymbol{b}_2'\cdot\boldsymbol{\rho})+\cos\big((\boldsymbol{b}_1'+\boldsymbol{b}_2')\cdot\boldsymbol{\rho}\big)\big], \tag{19}$$

here $\boldsymbol{b}_1'$, $\boldsymbol{b}_2'$ are rotated by $30°$ relative to $\boldsymbol{b}_1$ and $\boldsymbol{b}_2$; therefore, we have artificially constructed the potential of a quasicrystal surface corresponding to a bilayer graphene twisted by $30°$.

### *2.1.4 Water Molecule Adsorption on Graphene*

Having established the analytical expression for the interaction between a single atom and an infinite graphene substrate [Eq. (11)], we now extend its applicability to molecules with internal structures through a rigid-body approximation. Within this framework, the molecular orientation is characterized by its normal vector, the atoms of the molecule are treated individually, and the total adsorption energy is obtained by summing the interaction energies of its constituent atoms. While this approximation neglects intramolecular deformation, it is suitable for rigid molecules such as water, where internal degrees of freedom are weakly coupled to the adsorption process.

Water adsorption on graphene tests the treatment of molecular orientation, using the computational protocol of [40]. In that work, reference DFT calculations were performed to obtain binding energy (BE) curves, sliding potential energy surfaces (PESs), and rotational energy profiles for rotation about different axes ($x$, $y$, and $z$) at a fixed height above the graphene plane. The KC parameters are fitted to these DFT reference data, thereby capturing the essential registry-dependent interactions between the water molecule and the graphene surface. The fitted KC potential is then used in conjunction with our analytical expression to compute the energy variation as a function of rotation angle.

Molecular orientation enters through the two projection distances in the registry-dependent potential:

$$\boldsymbol{\rho}^2_{\mathrm{sub}\to\mathrm{ad}} = \boldsymbol{r}^2_{\mathrm{sub-ad}} - (\boldsymbol{r}_{\mathrm{sub-ad}}\cdot\boldsymbol{n}_{\mathrm{sub}})^2, \qquad \boldsymbol{\rho}^2_{\mathrm{ad}\to\mathrm{sub}} = \boldsymbol{r}^2_{\mathrm{sub-ad}} - (\boldsymbol{r}_{\mathrm{sub-ad}}\cdot\boldsymbol{n}_{\mathrm{ad}})^2. \tag{20}$$

Vector $\boldsymbol{r}_{\mathrm{sub-ad}}$ represents the relative position vector between the substrate atom and the adsorbed atom, and together with the orientations of the corresponding atoms, it is used to calculate the projection distances to obtain $\boldsymbol{\rho}^2_{\mathrm{sub}\to\mathrm{ad}}$ and $\boldsymbol{\rho}^2_{\mathrm{ad}\to\mathrm{sub}}$. Here, our introduction of the normal vector is consistent with the definition in [43], except that here we explicitly distinguish between the orientation of the substrate atoms $\boldsymbol{n}_{\mathrm{sub}}$ and the orientation of the adsorbed atoms $\boldsymbol{n}_{\mathrm{ad}}$. In our previous calculations, the orientations of both the adsorbate atom (denoted as ad) and the substrate atom (denoted as sub) were set to $(0,0,1)$ by default. In this case, the two projection distances both correspond to the projected distance in three-dimensional space, namely $\rho^2 = r^2 - z^2$. By keeping the substrate atom orientation fixed at $(0,0,1)$, the discrete translational symmetry Eq. (1) remains satisfied. Therefore, the Fourier series approach is still applicable. Orientation therefore enters through the potential integral defining the Fourier coefficients in Eq. (6). In the original integral, the integration variable $\boldsymbol{X}$ corresponds to the projected position in space. Upon introducing the molecular orientation, the term $\boldsymbol{\rho}^2_{\mathrm{ad}\to\mathrm{sub}}$ appearing in the potential energy couples the spatial projected position with the components $(n_x, n_y, n_z)$ of the normal vector $\boldsymbol{n}_{\mathrm{ad}}$. SI A.2 derives the Fourier coefficients with this normal-vector dependence. The key step consists of decomposing the original potential function into an isotropic part and an anisotropic part. The resulting total energy expression, truncated at the first order of the Fourier expansion, is given by:

$$u(\boldsymbol{\rho}, z; \boldsymbol{n}_{\mathrm{ad}}) = \frac{2}{a_s}\left(V_0^{\mathrm{I}} + V_0^{\mathrm{A}}\right) - \frac{2}{a_s}\sum_{i=1}^{3}\left[\left(V_1^{\mathrm{I}} + \mathrm{Re}\left(V_1^{\mathrm{A}}\right)\right)\cos(\boldsymbol{b}_i\cdot\boldsymbol{\rho}) - \mathrm{Im}\left(V_1^{\mathrm{A}}\right)\sin(\boldsymbol{b}_i\cdot\boldsymbol{\rho})\right], \tag{21}$$

while the expressions for the isotropic part $V^{\mathrm{I}}$ and the anisotropic part $V^{\mathrm{A}}(\boldsymbol{b}_i)$ for KC potential are given by

$$V_{\boldsymbol{n}}^{\mathrm{I}} = 2\pi\int_0^{\infty} \rho' J_0(|\boldsymbol{G}|\rho') u_{\mathrm{KC}}^{\mathrm{I}}(\rho', z)\mathrm{d}\rho', \tag{22}$$

$$V_{\boldsymbol{n}}^{\mathrm{A}} = \iint_{\infty} \mathrm{d}x'\mathrm{d}y' e^{-i\boldsymbol{G}\cdot(x',y')} e^{-\lambda\left(\sqrt{x'^2+y'^2+z^2}-z_0\right)-\frac{\xi'^2}{\delta^2}}\left[C_0 + C_2\left(\frac{\xi'}{\delta}\right)^2 + C_4\left(\frac{\xi'}{\delta}\right)^4\right]. \tag{23}$$

Following the same convention as above, the subscript $V_0^{\mathrm{I}}$, $V_0^{\mathrm{A}}$ in the potential energy integration result corresponds to the case where the reciprocal lattice vector vanishes, $\boldsymbol{G} = \boldsymbol{0}$, while the subscript 1 corresponds to the first-order reciprocal lattice vectors. Here the isotropic part in potential $u_{\mathrm{KC}}^{\mathrm{I}}$ is similar to Eq. (8) but with the $\boldsymbol{\rho}^2_{\mathrm{ad}\to\mathrm{sub}}$ component removed

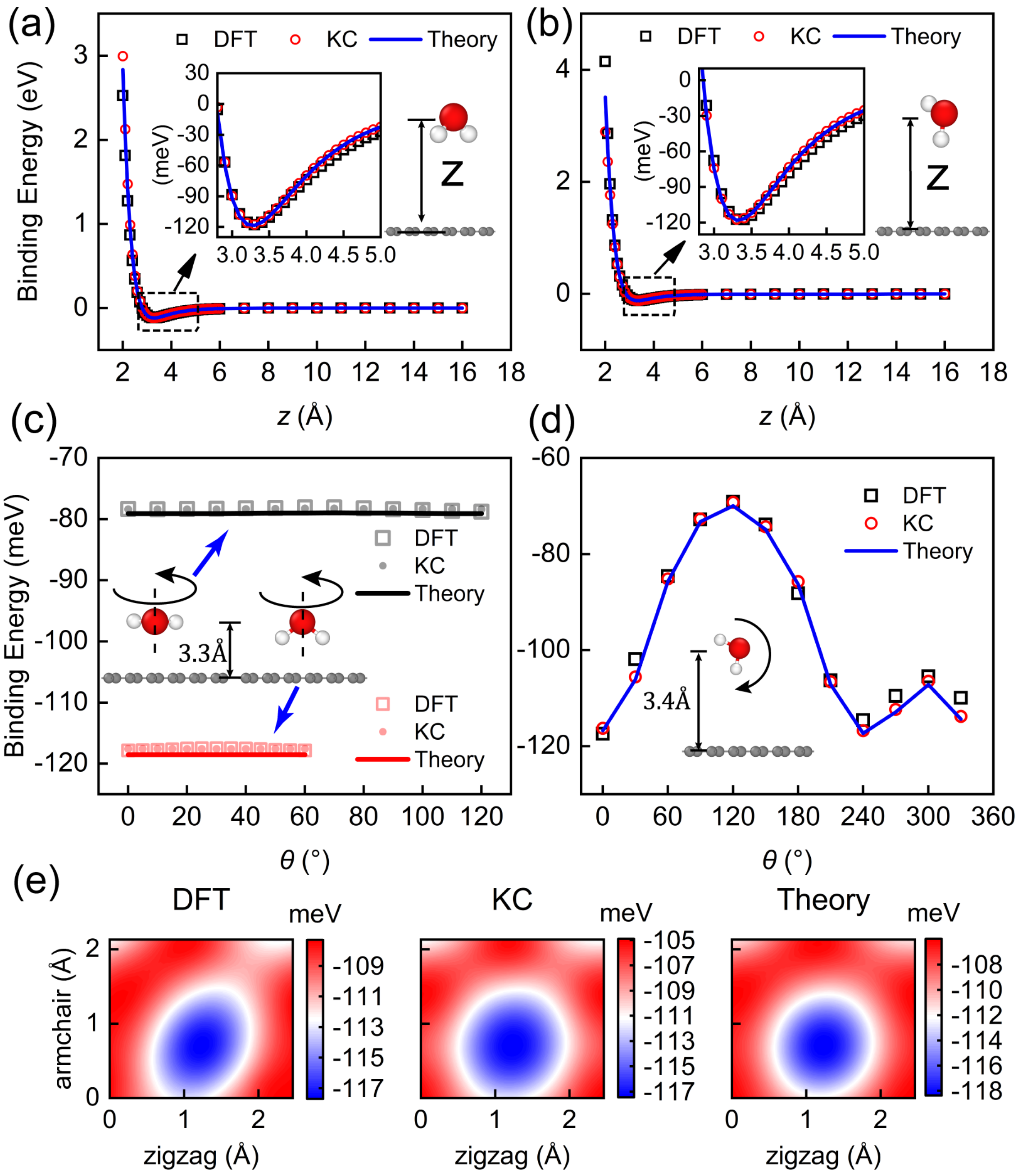

Fig. 3. Binding energy and potential energy surface of a water molecule interacting with graphene substrate. The graphene substrate serves as the rigid infinite substrate in the theoretical model. The three distinct rotation modes of the water molecule are illustrated in the figure.

$$u^{\mathrm{I}}_{\mathrm{KC}}(\rho,z)=\left\{c+\left[c_0+c_2\left(\frac{\rho}{\delta}\right)^2+c_4\left(\frac{\rho}{\delta}\right)^4\right]e^{-\left(\frac{\rho}{\delta}\right)^2}\right\}e^{-\lambda\left(\sqrt{\rho^2+z^2}-z_0\right)}-\frac{Az_0^6}{(\rho^2+z^2)^3}. \tag{24}$$

The anisotropy in the energy is governed by the anisotropic parameter $\xi'=\sqrt{(x'^2+y'^2+z^2)-\left(x'n_x+y'n_y+zn_z\right)^2}$ which incorporates the components of the normal vector. $\mathrm{Re}(\cdot)$ and $\mathrm{Im}(\cdot)$ denotes the real and imaginary parts of the integral, respectively.

Fig. 3 compares the analytical binding-energy curves, sliding potential energy surfaces and rotational profiles with the reference DFT data and direct calculations using the fitted KC potential.

Agreement across these three types of motion shows that the Fourier representation preserves both registry and orientation dependence in the molecular adsorption landscape.

A two-dimensional ice crystal on graphene extends the molecular calculation to a finite molecular lattice. The lattice geometry is taken from the water islands in [41], for which one ice lattice vector is commensurate with four graphene lattice vectors. Because the graphene lattice constant of that reference configuration (2.460 Å) is very close to, but not identical with the value we adopted (2.463 Å) in the water molecule calculation, so that the resulting interface is near-commensurate rather than exactly commensurate. We construct a circular ice flake of radius $R = 100$ Å from this configuration, placing its center at the center of the hexagonal ice bilayer, with the ice lattice aligned to the substrate ($\theta = 0°$). Taking the average height of the oxygen atoms of the lower water layer as the reference plane, and varying this height from 2 to 8 Å, give the binding energy curve in Fig. 4. With the reference height fixed at 3.5 Å, sliding the flake yield the potential energy surface at $\theta = 0°$ and $\theta = 4°$.

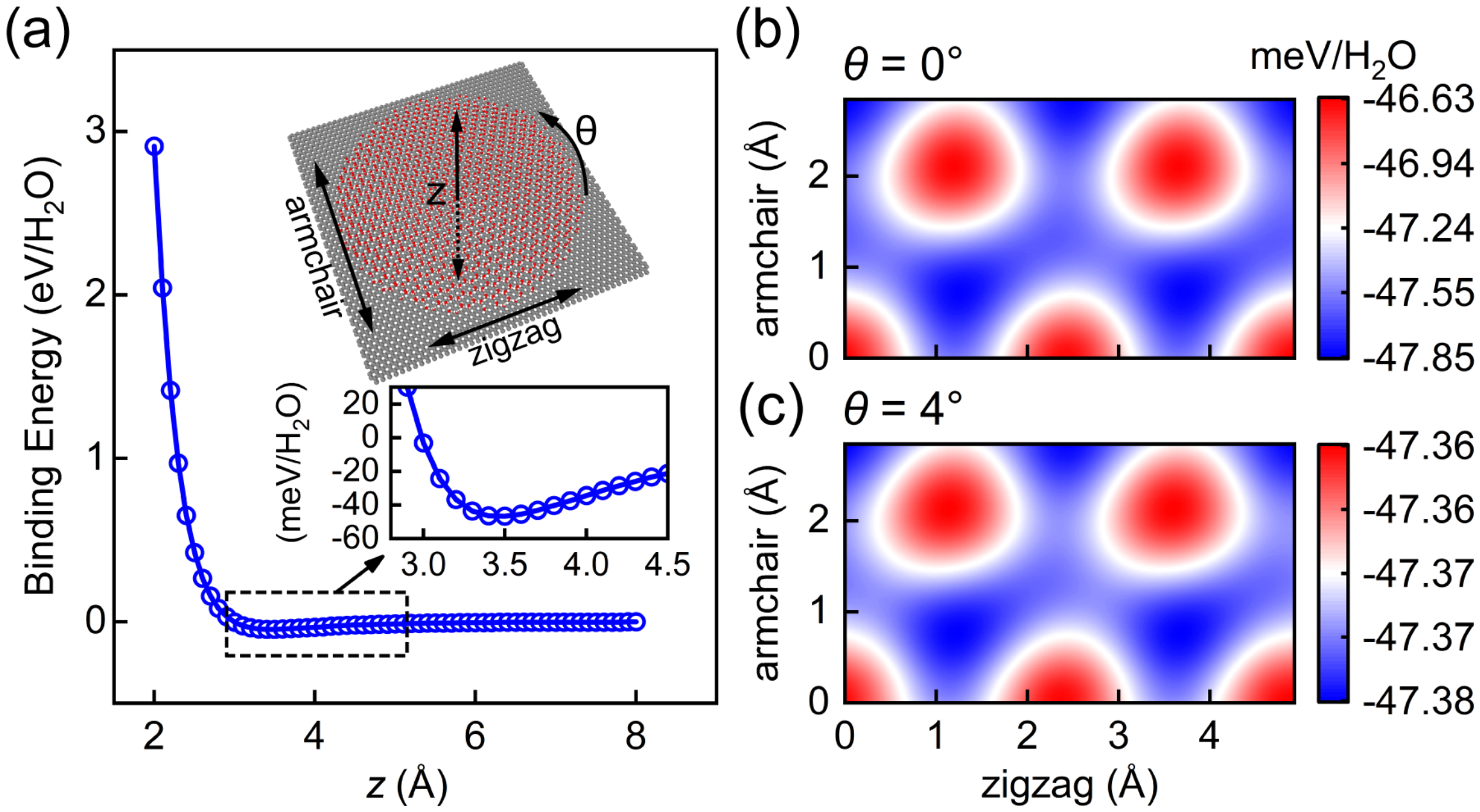

Fig. 4. Interaction of a 2D ice lattice with graphene substrate. (Left) Binding energy per water molecule as a function of the interlayer separation $z$ for a circular ice flake of radius 100 Å on graphene at $\theta = 0°$; (Right) Sliding potential energy surface per water molecule over substrate along the zigzag and armchair direction at $\theta = 0°$ and $\theta = 4°$. The corrugation is about 1.2 meV/$H_2O$ at $\theta = 0°$ but reduces to only 0.02 meV/$H_2O$ at $\theta = 4°$.

The resulting energy corrugation exhibits a pronounced registry dependence on the relative alignment between the ice lattice and the underlying graphene sheet. Twisting the flake away from the near commensurate orientation to $4°$ reduces the corrugation significantly. Specifically, this twist lowers the barrier by a factor of 60, from 1.2 to 0.02 meV/$H_2O$. This demonstrates that geometric control can drive the interface from near-commensurability towards incommensurability and suppress the barrier. The comparison with [41] remains qualitative. The potential parameters used here were parameterized against the many-body dispersion corrected DFT data for water-graphene interface [40], whereas the DFT in [41] were performed on water islands adsorbed on Cu-supported graphene, so the absolute energies are not directly comparable. These calculations connect the orientation-dependent molecular interaction to the registry-dependent energy of a finite ice lattice, providing the basis for the boundary-scaling tests below.

## 2.2 Interlayer Energetics of Twisted Bilayer Graphene

The energy-density representation connects molecular adsorption and crystalline contacts by combining the substrate potential with the atomic density of the upper layer. For twisted bilayer graphene, the infinite-area limit determines the bulk energy, while finite integration domains retain the contribution of contact boundaries. We consider two infinite graphene sheets with a relative twist angle, which provides the foundation for understanding the periodic moiré superlattice and serves as the bulk limit for larger systems. Then we address the finite-size case, where one layer are truncated to form finite graphene flakes placed on an infinite substrate. This geometry introduces boundary effects and edge contributions that are absent in the infinite limit and are essential for understanding the size-dependent behavior of realistic contacts.

This common representation makes lattice registry and finite geometry explicit in the energy, forming the basis for subsequent analysis of mechanical, thermal, and tribological properties of twisted vdW heterostructures.

### *2.2.1 Energy Distribution Density Formalism*

The central step is to express the upper lattice as an atomic density, so that its interaction with the substrate becomes a spatial integral. For two rigid, periodic 2D layers, summing Eq. (4) over the upper-layer atoms gives the total interaction energy. To retain a spatial variable analogous to $\boldsymbol{\rho}$ in Eq. (2), we represent the upper-layer atomic positions by a density distribution of Dirac delta functions. Multiplying this density by the substrate potential preserves the lattice information and enables the energy to be integrated over the contact domain.

The atomic density distribution ($f_\rho$) is defined as follows:

$$f_\rho(\boldsymbol{\tau};\boldsymbol{T}) = \sum_{l'}\sum_{k'} \delta\big(\boldsymbol{\tau} - (\boldsymbol{L}' + \boldsymbol{B}'_{k'} + \boldsymbol{T})\big) \tag{25}$$

where superscripts denote physical quantities associated with the upper lattice. The vector $\boldsymbol{L}'$ represents a linear combination of lattice vector $\boldsymbol{a}'_1$, $\boldsymbol{a}'_2$ of upper layer, and $\boldsymbol{B}'_{k'}$ denotes the coordinates of the $k'$-th atom within a unit cell, as shown in Fig. 6 (a). Here, $\boldsymbol{T}$ denotes the translation vector of the upper lattice relative to the substrate, while the vector $\boldsymbol{\tau}$ serves as a probe variable to capture the periodicity of atomic distribution for the upper lattice.

The atomic density function Eq. (25) possesses dimensions of inverse area (i.e., areal density) and serves as the basis for calculating the total potential energy. Multiplying this density function by the single-atom potential and an infinitesimal area element $\mathrm{d}\boldsymbol{\tau}$ yields $f_\rho(\boldsymbol{\tau};\boldsymbol{T})\mathrm{d}\boldsymbol{\tau}$, which represents the number of atoms in the area element at position $\boldsymbol{\tau}$. The product $U_\rho \mathrm{d}\boldsymbol{\tau} = u(\boldsymbol{\tau}, z) f_\rho(\boldsymbol{\tau};\boldsymbol{T})\mathrm{d}\boldsymbol{\tau}$ then gives the potential energy contribution from this infinitesimal region, where $U_\rho$ is defined as the interlayer energy distribution density. Consequently, the total potential energy of an arbitrary crystalline layer above an infinite substrate is obtained by integrating the energy density over the entire area of the upper lattice:

$$U = \iint_{A'} U_\rho \mathrm{d}\boldsymbol{\tau} = \iint_{A'} u(\boldsymbol{\tau}, z) f_\rho(\boldsymbol{\tau}; \boldsymbol{T}) \mathrm{d}\boldsymbol{\tau}. \tag{26}$$

In analogy with the Fourier expansion of the single-atom potential (Eq. (2)), we now expand the atomic density distribution of the upper layer with respect to its own lattice periodicity: $f_\rho(\boldsymbol{\tau}; \boldsymbol{T}) = \sum_{\boldsymbol{n}'} C'_{\boldsymbol{n}'} e^{i\boldsymbol{G}' \cdot (\boldsymbol{\tau} - \boldsymbol{T})}$, and with the expansion we used in Eq. (2): $u(\boldsymbol{\rho}, z) = \sum_{\boldsymbol{n}} C_{\boldsymbol{n}} e^{i\boldsymbol{G} \cdot \boldsymbol{\rho}}$, the energy density distribution can be written in the form of a double Fourier expansion

$$U_\rho = \sum_{\boldsymbol{n}} \sum_{\boldsymbol{n}'} C_{\boldsymbol{n}} C'_{\boldsymbol{n}'} e^{i\boldsymbol{G} \cdot \boldsymbol{\tau}} e^{i\boldsymbol{G}' \cdot (\boldsymbol{\tau} - \boldsymbol{T})}, \tag{27}$$

the new Fourier coefficient $C'_{\boldsymbol{n}'}$:

$$C'_{\boldsymbol{n}'} = \frac{1}{a'_s} \left[ \sum_{k'} \exp\left(-i\boldsymbol{G}' \cdot \boldsymbol{B}'_{k'}\right) \right], \tag{28}$$

An integral analogous to Eq. (6) appears, $\int_{S'_\infty} \delta(\boldsymbol{X}) \exp(-i\boldsymbol{G}' \cdot \boldsymbol{X}) \, \mathrm{d}\boldsymbol{X}$, but due to the Dirac delta function in the integrand, it yields a value of 1.

Two points require clarification. First, the translation vector $\boldsymbol{T}$ is introduced to represent the relative displacement of the upper lattice. Consequently, within the global coordinate system fixed on the substrate, the periodic variable for the upper lattice becomes $\boldsymbol{\tau} - \boldsymbol{T}$. Second, to employ a Fourier series expansion, the upper lattice must be periodically extended to infinity, making both layers infinite which is shown in Fig. 6 (b). However, since the atomic density is localized, the energy of a finite-sized flake can still be computed by integrating only over its actual area.

### 2.2.2 *Energetics of Infinite Twisted Bilayer*

Before we calculate the interlayer interaction between two infinitely large graphene layers, the three important points should be noted:

(i) Our theoretical model assumes an infinite substrate. If the upper lattice is also infinite, the total energy will diverge. Therefore, the energy must be normalized per unit area,

$$w = \frac{1}{A'} \iint_{A'} U_\rho \mathrm{d}\boldsymbol{\tau}, \tag{29}$$

(ii) With this normalization, the interaction energy of infinite twisted bilayer graphene is obtained by taking the integration domain in Eq. (26) to infinity. For $\boldsymbol{G} + \boldsymbol{G}' \neq \boldsymbol{0}$,

$$\frac{1}{A'} \left| \iint_{A'} e^{i(\boldsymbol{G}+\boldsymbol{G}') \cdot \boldsymbol{\tau}} \mathrm{d}\boldsymbol{\tau} \right| \leq \frac{1}{A' |\boldsymbol{G} + \boldsymbol{G}'|} \oint_{\partial A'} \left| (\widehat{\boldsymbol{G}} + \widehat{\boldsymbol{G}}') \cdot \widehat{\boldsymbol{n}} \right| \mathrm{d}l \propto \frac{R}{R^2} = \frac{1}{R},$$

the rapidly oscillating phase factors in the integrand average to zero in the infinite-domain limit.

(iii) The constant term retained in the infinite-area limit arises from two sources: (a) the Fourier component with $(n_1, n_2) = (n'_1, n'_2) = (0,0)$ in Eq. (27), and (b) the $\boldsymbol{G} + \boldsymbol{G}' = \boldsymbol{0}$ mode that emerges specifically in commensurate state.

In case (iii. a), only the constant term remains in the double Fourier form of the energy distribution density,

$$U_\rho = \frac{4V_0(z)}{a_s^2}, \tag{30}$$

This term is the interaction energy density obtained by averaging out the lateral registry of both layers.

In case (iii. b), which naturally includes the first case, the form of the potential energy density distribution for twisted bilayer graphene can be simplified as follows:

$$U_\rho = \frac{1}{a_s^2}\sum_{\boldsymbol{n}}\sum_{\boldsymbol{n}'} V_{\boldsymbol{n}}(z)\left[\sum_k \exp(-i\boldsymbol{G}\cdot\boldsymbol{B}_k)\right]\left[\sum_k \exp(i\boldsymbol{G}\cdot\boldsymbol{B}_k)\right] e^{i\boldsymbol{G}\cdot\boldsymbol{T}}, \tag{31}$$

Here we substitute the complete Fourier coefficients Eq. (5) and Eq. (28) into Eq. (27). Taking the first-order approximation (see SI A.3) yields

$$U_{\rho 1} = \frac{4V_0(z)}{a_s^2} + \frac{2V_1(z)}{a_s^2}P_T(\boldsymbol{T}). \tag{32}$$

Here $P_T(\boldsymbol{T}) = \cos(\boldsymbol{b}_1\cdot\boldsymbol{T}) + \cos(\boldsymbol{b}_2\cdot\boldsymbol{T}) + \cos\big((\boldsymbol{b}_1+\boldsymbol{b}_2)\cdot\boldsymbol{T}\big)$ is defined as a translation term. And the energy for the commensurate configuration:

$$w_1 = \frac{1}{A'}\iint_{A'} U_{\rho 1}\mathrm{d}\boldsymbol{\tau} = \frac{4V_0(z)}{a_s^2} + \frac{2V_1(z)}{a_s^2}P_T(\boldsymbol{T}), \tag{33}$$

and for the incommensurate case, due to the reasons outlined in (ii), it suffices to retain only the constant Fourier term.

$$w_2 = \frac{4V_0(z)}{a_s^2}. \tag{34}$$

It can be clearly seen that, in the commensurate state (Eq. (33)), the form of the interaction energy is consistent with that of a single atom Eq. (11). In deriving the bilayer model, the in-plane origins of the upper and lower graphene layers are placed at the centers of the hexagonal lattices by choosing $B_1 = B_1' = \left(\frac{1}{3},\frac{1}{3}\right)$ and $B_2 = B_2' = \left(\frac{2}{3},\frac{2}{3}\right)$, so that the local stacking configuration in our theory is AA stacking. It can be readily verified that by using the single-atom theory, we place two atoms at the AA stacking sites, respectively:

$$\begin{aligned} u_{\text{local-AA}} &= u\left(\boldsymbol{\rho} + \frac{1}{3}(\boldsymbol{a}_1+\boldsymbol{a}_2), z\right) + u\left(\boldsymbol{\rho} - \frac{1}{3}(\boldsymbol{a}_1+\boldsymbol{a}_2), z\right) \\ &= \frac{4V_0(z)}{a_s} + \frac{2V_1(z)}{a_s}\left[\cos(\boldsymbol{b}_1\cdot\boldsymbol{\rho}) + \cos(\boldsymbol{b}_2\cdot\boldsymbol{\rho}) + \cos\big((\boldsymbol{b}_1+\boldsymbol{b}_2)\cdot\boldsymbol{\rho}\big)\right] \end{aligned} \tag{35}$$

and then averaging the energy over the area of a Bravais lattice $a_s$, we recover the commensurate energy per unit area of Eq. (33), which confirms the consistency between the single-atom picture and the density-based bilayer formulation.

## 2.3 Energetics of finite flakes on infinite substrates

### *2.3.1 Circular Flake*

Based on Eq. (26)and (27), the general interaction expression for a circular flake can be obtained by directly integrating over a circular domain of radius $R$ centered at $\boldsymbol{T}$:

$$U = \int_0^R\int_0^{2\pi}\sum_{\boldsymbol{n}}\sum_{\boldsymbol{n}'} C_{\boldsymbol{n}}C'_{\boldsymbol{n}'}e^{i(\boldsymbol{G}+\boldsymbol{G}')\cdot(\boldsymbol{\tau}'+\boldsymbol{T})}e^{-i\boldsymbol{G}'\cdot\boldsymbol{T}}\,d\boldsymbol{\tau}' = \sum_{\boldsymbol{n}}\sum_{\boldsymbol{n}'} C_{\boldsymbol{n}}C'_{\boldsymbol{n}'}e^{i\boldsymbol{G}\cdot\boldsymbol{T}}\int_0^R\int_0^{2\pi} e^{i(\boldsymbol{G}+\boldsymbol{G}')\cdot\boldsymbol{\tau}'}\,d\boldsymbol{\tau}'$$

$$= 2\pi R^2 \sum_{\boldsymbol{n}} \sum_{\boldsymbol{n}'} C_{\boldsymbol{n}} C'_{\boldsymbol{n}'} \exp(i\boldsymbol{G} \cdot \boldsymbol{T}) \frac{J_1(|\boldsymbol{G} + \boldsymbol{G}'|R)}{|\boldsymbol{G} + \boldsymbol{G}'|R}, \quad (36)$$

here, we use a change of variables $\boldsymbol{\tau}' = \boldsymbol{\tau} - \boldsymbol{T}$ to shift the center of the circular integration region back to the origin, where $J_1$ is the first-order Bessel function of the first kind. In the expression $\boldsymbol{G} + \boldsymbol{G}'$ corresponds to the moiré pattern [1] formed between the layers and $\boldsymbol{G}$ denotes the reciprocal lattice vector of the substrate. The moiré reciprocal lattice vector appearing in the argument of the Bessel function is independent of the sliding position $\boldsymbol{T}$ of the circular flake. This implies that, for a circular flake, the landscape of the sliding potential energy surface depends entirely on the lattice of the substrate, namely, the discrete translational symmetry Eq. (1) imposed by the substrate lattice, while the moiré pattern only modulates the magnitude of the energy barrier. For the commensurate AB stacking in BLG system, this substrate-imposed landscape is quasi-isotropic: the sixfold symmetry of the graphite lattice renders the interlayer shear response a single elastic modulus [46], and makes the static friction nearly independent of the sliding direction when the registry coupling is strong [48]. Introducing a twist angle departs from this limiting case, because the moiré term then modulates the barrier magnitude and thereby diversifies the sliding behavior of graphene. The magnitude of the energy barrier, in turn, governs the static friction force.

For the circular flake described by the Bessel function, the previous work [14] has already demonstrated that the static friction force exhibits a scaling law of $R^{0.5}$, i.e., $A^{0.25}$(where $A$ is the area). The appearance of the Bessel function in our formulation is a consequence of the circular boundary. Moreover, the summation over reciprocal lattice vectors does not change with the flake size; the number of terms retained in the summation only affects the pre-factor of $R$, not the scaling exponent itself. Therefore, our theoretical analysis suggests that the $A^{0.25}$ scaling law of static friction for circular flakes may be universal across different lattice types.

A circular two-dimensional ice flake tests this boundary-scaling prediction for a molecular lattice, with its total energy calculated from the water-molecule expression in Eq. (21). For each radius, the ice lattice is placed at a prescribed twist angle, the flake is slid, and the barrier amplitude $\Delta U$ is extracted from the resulting potential energy surface; twist angles of 0° and 4° are considered. The resulting scaling of $\Delta U$ with the flake area is shown in log-log scale in Fig. 5. At 4° the amplitude envelope follows $\Delta U \propto A^{0.25}$ over the whole range of sizes, in agreement with the analysis above. At 0° the amplitude instead grows linearly with the area, $\Delta U \propto A^1$, as expected when moiré reciprocal lattice reduces to $\boldsymbol{0}$ in Eq. (36), namely in commensurate state, the Bessel function term reduces to non-zero constant and the only R-dependent term is area $\pi R^2$. The amplitude follows this linear law closely over most of the size range and lies slightly below it at the largest sizes. This small deviation has a clear origin: the configuration adopted here is near-commensurate, one ice lattice vector differs slightly from the graphene substrate, not exactly commensurate as assumed in theory. The residual mismatch then acts as an effective finite twist, so that a linear-in-area growth cannot persist indefinitely. Recovering both the $A^{0.25}$ law for the incommensurate contacts and the $A^1$ law near commensurability for a lattice entirely different from graphene confirms that the scaling exponent is dictated by the boundary geometry and the registry state.

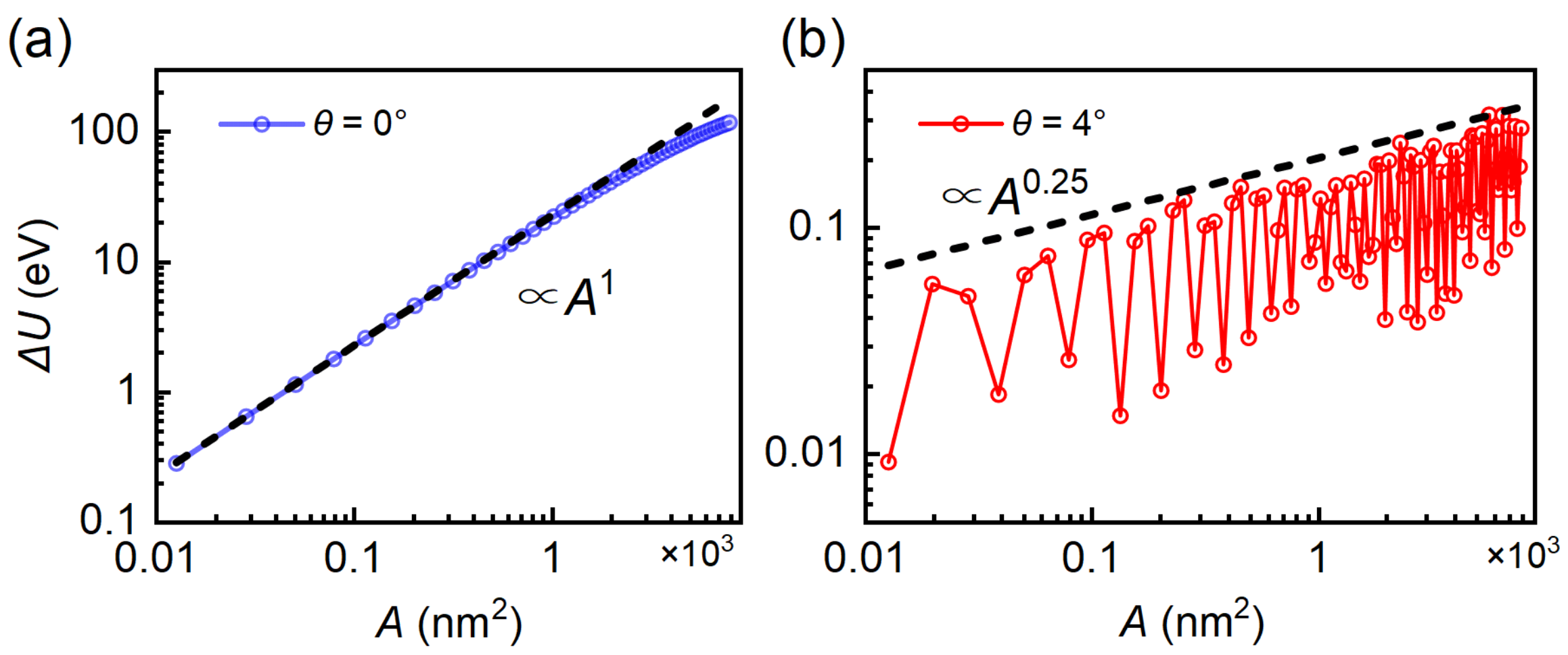

Fig. 5. Size scaling of the barrier amplitude for circular 2D ice flakes. Log-log plot of the sliding barrier amplitude $\Delta U$ against the flake area $A$ at $\theta = 0°$ (left) and $\theta = 4°$ (right), obtained by varying the flake radius at fixed twist angle. At $\theta = 0°$ the amplitude closely follows $\Delta U \propto A^1$ over most of the size range, departing slightly below the linear law only at the largest sizes: the configuration is near-commensurate rather than exactly commensurate. At $\theta = 4°$ the amplitude oscillates strongly with size and its envelope follows $\Delta U \propto A^{0.25}$.

For twisted bilayer graphene (tBLG), the general expression simplifies under the leading moiré-mode approximation. By truncating the double Fourier series in reciprocal space to include only the origin and the shortest reciprocal lattice vectors generated by the two layers, we arrive at an expression for the interlayer energy density (see SI A.3 for derivation details)

$$U_\rho(z, \theta, \boldsymbol{T}, \boldsymbol{\tau}) = \frac{4V_0(z)}{a_s^2} + \frac{2V_1(z)}{a_s^2} P_\mathrm{m}(\boldsymbol{\tau}; \theta, \boldsymbol{T}), \tag{37}$$

with moiré pattern term $P_\mathrm{m}$ is given by:

$$\begin{aligned} P_\mathrm{m}(\boldsymbol{\tau}; \theta, \boldsymbol{T}) &= \cos\left(\frac{4\pi}{\sqrt{3}a_\mathrm{m}} \hat{\boldsymbol{e}}\left(\frac{\theta}{2}\right) \cdot \boldsymbol{\tau} + \frac{4\pi}{\sqrt{3}a} \hat{\boldsymbol{e}}\left(\theta + \frac{\pi}{2}\right) \cdot \boldsymbol{T}\right) \\ &+ \cos\left(\frac{4\pi}{\sqrt{3}a_\mathrm{m}} \hat{\boldsymbol{e}}\left(\frac{\theta}{2} - \frac{\pi}{3}\right) \cdot \boldsymbol{\tau} + \frac{4\pi}{\sqrt{3}a} \hat{\boldsymbol{e}}\left(\theta - \frac{\pi}{3} + \frac{\pi}{2}\right) \cdot \boldsymbol{T}\right) \\ &+ \cos\left(\frac{4\pi}{\sqrt{3}a_\mathrm{m}} \hat{\boldsymbol{e}}\left(\frac{\theta}{2} + \frac{\pi}{3}\right) \cdot \boldsymbol{\tau} + \frac{4\pi}{\sqrt{3}a} \hat{\boldsymbol{e}}\left(\theta + \frac{\pi}{3} + \frac{\pi}{2}\right) \cdot \boldsymbol{T}\right). \end{aligned} \tag{38}$$

Here, $a$ denotes the lattice constant of graphene, and $a_\mathrm{m} = \frac{a}{2\sin(\theta/2)}$ corresponds to the moiré superlattice period [1,12,15]. The unit vector $\hat{\boldsymbol{e}}(\theta)$ is defined as $\hat{\boldsymbol{e}}(\theta) = (\cos\theta, \sin\theta)$. Rotating the coordinate system by $\theta/2$ around the positive $z$-axis transforms the expression into the widely-used approximate form for the energy density of bilayer graphene: $2\cos\left(\frac{2\pi x}{\sqrt{3}a_\mathrm{m}}\right)\cos\left(\frac{2\pi y}{a_\mathrm{m}}\right) + \cos\left(\frac{4\pi x}{\sqrt{3}a_\mathrm{m}}\right)$ (see SI A.3). [15,39,49,50]. When $\theta$ increases from $0°$ to beyond $30°$, the originally selected shortest reciprocal lattice vector is replaced by the next-shortest one, rendering the original formula inapplicable. For twist angles outside the range of $\theta \in [0, \pi/6]$, the formula requires selecting the shortest reciprocal lattice vector that can be formed at the given angle. Details are provided in the SI

3.

With the energy density established in Eq. (37), the total energy for a circular graphene flake of radius $R$ is obtained by integrating Eq. (26) over a circular domain centered at position $\boldsymbol{T}$

$$U_{\mathrm{c}}=\iint_{A'} U_{\rho}\mathrm{d}\boldsymbol{\tau}=\frac{4\pi R^2 V_0(z)}{a_s^2}+\frac{2V_1(z)}{a_s^2}\int_0^{2\pi}\int_0^R P_{\mathrm{m}}(\boldsymbol{\tau}'+\boldsymbol{T};\theta,\boldsymbol{T})\mathrm{d}\boldsymbol{\tau}', \tag{39}$$

the variable substitution $\boldsymbol{\tau}'=\boldsymbol{\tau}-\boldsymbol{T}$ is implemented again through a translation of the circle's center back to the origin

$$U_{\mathrm{c}}(z,\theta,\boldsymbol{T},R)=\frac{4\pi R^2 V_0(z)}{a_s^2}+\frac{2V_1(z)}{a_s^2}P_T(\boldsymbol{T})\underbrace{\left[2\pi R^2\frac{J_1\left(\frac{4\pi R}{\sqrt{3}a_{\mathrm{m}}}\right)}{\frac{4\pi R}{\sqrt{3}a_{\mathrm{m}}}}\right]}_{\equiv\Phi_{\mathrm{c}}(R,a_{\mathrm{m}})}$$

$$=\frac{4\pi R^2 V_0(z)}{a_s^2}+\frac{2V_1(z)}{a_s^2}P_T(\boldsymbol{T})\Phi_{\mathrm{c}}(R,a_{\mathrm{m}}) \tag{40}$$

with a translation term $P_T(\boldsymbol{T})=\cos(\boldsymbol{b}_1\cdot\boldsymbol{T})+\cos(\boldsymbol{b}_2\cdot\boldsymbol{T})+\cos\big((\boldsymbol{b}_1+\boldsymbol{b}_2)\cdot\boldsymbol{T}\big)$ and a boundary term $\Phi_{\mathrm{c}}(R,a_{\mathrm{m}})$ that characterizes the geometric shape of the material.

To validate our formulation, we compare the energy computed from Eq. (40) with that obtained from direct atom-by-atom summation using Eq. (11). Fig. 6 presents a comprehensive validation of our analytical expression. Panel (b) demonstrates excellent agreement between our formula and atomistic calculations for a flake radius of 15 nm across a range of twist angles under both AA and AB stacking configurations. Panel (c-d) further confirms the accuracy of the predicted sliding energy surface for a circular flake with twist angle $\theta=4°$.

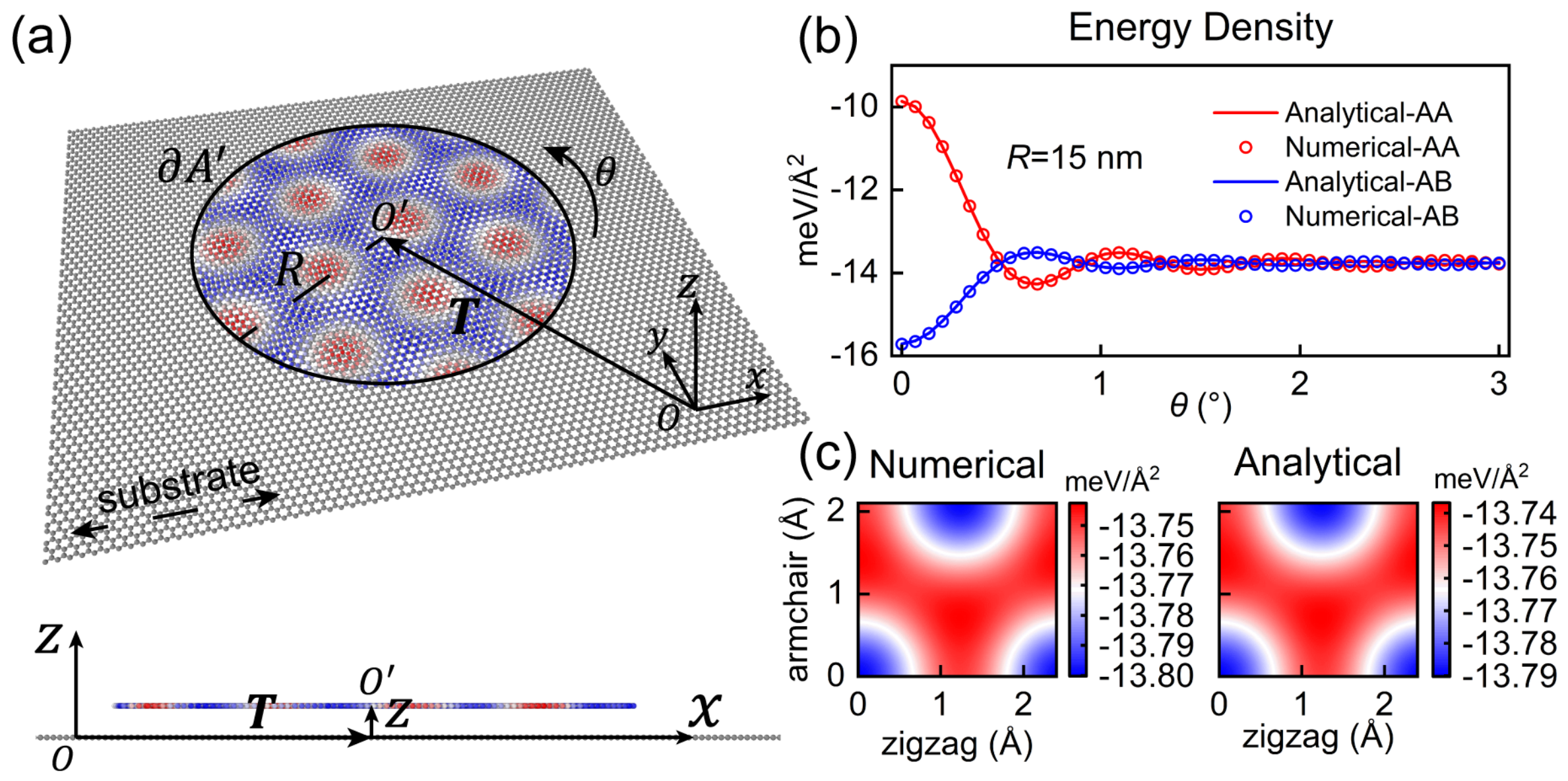

Fig. 6. Bilayer Graphene model and Circular Graphene Flakes. (a) Schematic of the bilayer graphene model. The local coordinate origin $O'$ of the upper layer is displaced by vector $\boldsymbol{T}$ and height $z$. The area occupied by the upper flake is denoted $A'$ and the interlayer twist angle is $\theta$. Physical quantities associated with the upper lattice are distinguished by a prime superscript; The Fourier expansion

requires the atomic density to be defined on an infinite domain $S'_\infty$, but the total energy is evaluated by integrating over the finite physical area $A'$ due to the localized nature of the atomic distribution; (b) Energy density variation of a circular graphene flake with a radius of 15 nm, rotated around the AA and AB stacking centers, computed using Eq. (40) and Eq. (11) with $z = 3.35$ Å; (c-d) Comparison of the sliding energy density surface for a circular flake computed by Eq. (40) and Eq. (11), for a flake radius of 15 nm and a twist angle of $\theta = 4°$.

[14] derived the circular-flake energy using a Fourier series. The present atomic-density formulation recovers this result while linking the finite-domain integral explicitly to the microscopic interaction coefficients. Equation (40) also establishes a direct microscopic link to the phenomenological continuum models written as $\pm\frac{2}{9}U_0\left[2\cos\left(2\pi x/\sqrt{3}a_\mathrm{m}\right)\cos(2\pi y/a_\mathrm{m})\right] + \cos\left(4\pi x/\sqrt{3}a_\mathrm{m}\right)$ with a fitted amplitude $U_0$. Since the bracketed function can be obtained by coordinates transformation of our $P_\mathrm{m}$ (SI A.3), comparison with Eq. (37) yields the microscopic identification

$$U_0 = \frac{9V_1(z)}{a_s^2}, \tag{41}$$

our framework derives the phenomenological corrugation amplitude from the KC potential with $V_1(3.35) = 17.8863\ \mathrm{meV}\cdot\text{Å}^2$ and $a_s = 5.248\ \text{Å}^2$, gives $U_0(3.35) = 5.845\ \mathrm{meV}/\text{Å}^2$, which is closed to value $5.85\ \mathrm{meV}/\text{Å}^2$ in [15]. Likewise, the sliding barrier of Eq. (40) scales as $R^2|J_1(x)/x|$ with $x = \frac{4\pi}{\sqrt{3}a_\mathrm{m}} \propto R\theta$, using the asymptotic envelope $|J_1(x)| \sim \sqrt{\frac{2}{\pi x}}$ reproduces the law $\Delta G \propto R^{\frac{1}{2}}\theta^{-\frac{3}{2}}$ in [14], with the zeros of $J_1(x)$ marking the sizes at which the barrier vanishes (complete moiré complementation).

A species-resolved atomic density extends the framework to interfaces with different lattice bases and atomic species. Just as we generalized the single-atom formulation of Eq. (7) to the general form of Eq. (14) in Sec. 2.1.2, the atomic distribution density can be similarly expanded to accommodate multiple atomic species. While Eq. (25) intrinsically accounts for different types of atoms, the interactions between distinct atom pairs must be treated explicitly. As defined previously, let $k$ denote the atoms in the substrate unit cell and $k'$ denote those in the upper-layer lattice. By summing the pairwise interactions over all relevant atomic combinations, Eq. (27) can be directly generalized as follows:

$$U_\rho(\boldsymbol{\tau};\boldsymbol{T}) = \frac{1}{a_s a'_s}\sum_{\boldsymbol{n},\boldsymbol{n}'}\left\{\sum_{k=1}^{N}\sum_{k'=1}^{N'} e^{-i\left(\boldsymbol{G}\cdot\boldsymbol{B}_k+\boldsymbol{G}'\cdot\boldsymbol{B}'_{k'}\right)} V_{\boldsymbol{n}}^{(k,k')}\left(z'_{k'}-z_k\right) e^{-i\boldsymbol{G}'\cdot\boldsymbol{T}} e^{i(\boldsymbol{G}+\boldsymbol{G}')\cdot\boldsymbol{\tau}}\right\} \tag{42}$$

where the species-resolved Fourier coefficient of the pair potential is

$$V_{\boldsymbol{n}}^{(k,k')}\left(\Delta z \overset{\mathrm{def}}{=} z'_{k'} - z_k\right) = \int_{S_\infty} u_{\mathrm{atom}}^{(k,k')}(\boldsymbol{X},\Delta z)\exp(-i\boldsymbol{G}\cdot\boldsymbol{X})\,\mathrm{d}\boldsymbol{X}. \tag{43}$$

Here, $a_s$ and $a'_s$ denote the areas of the Bravais lattices of the substrate and the upper lattice, respectively; $\boldsymbol{B}_k$ and $\boldsymbol{B}'_{k'}$ denote the local positions of the different atoms in the unit cells of the substrate and the upper lattice, respectively; $V_{\boldsymbol{n}}^{(k,k')}$ is the interaction potential term, determined by the interaction between atom $k$ in the substrate unit cell and atom $k'$ in the upper-lattice unit cell; $z_k$ and $z'_{k'}$ represent the respective atomic heights of the interacting atom pair; $\boldsymbol{G}$ and $\boldsymbol{G}'$ correspond to the reciprocal lattice vectors of the substrate and the upper lattice; the vector $\boldsymbol{T}$ correspond to the displacement of the upper lattice; the vector $\boldsymbol{\tau}$ is a point of interest in space and $U_\rho(\boldsymbol{\tau};\boldsymbol{T})$ corresponds to the potential energy density at position $\boldsymbol{\tau}$. As expressed in Eq. (42), the species-resolved expression retains both the sliding-induced phase shift and the moiré pattern. Its Fourier coefficients encode the chosen pair interactions and lattice bases, making the contributions from different atom pairs explicit. Furthermore, the framework is highly versatile and extends naturally to various heterojunctions. For instance, it can accurately describe systems such as graphene paired with $h$-BN, $MoS_2$, or gold (in both flake-on-substrate and substrate-on-flake configurations). Overall, this demonstrates the broad applicability of our model across diverse van der Waals interfaces.

### *2.3.2 Regular Polygonal Flakes*

For a regular polygon, the domain integral makes edge orientation an explicit variable in the energy. The polygon is uniquely defined by the position of its center $\boldsymbol{T}$, the number of sides $N$, and the circumradius $R$, as illustrated in Fig. 7(a). Additionally, an in-plane orientation angle $\beta$ is introduced, defined such that $\beta = 0$ when one edge of the polygon is perpendicular to the *x*-axis. Fig. 7(b) shows triangular flakes with identical stacking but different edge orientations ($\Delta\beta = \pi/6$), highlighting the role of orientation in determining the interfacial energetics.

Analogous to the circular case, the total energy decomposes into a translation-dependent term and boundary terms that encode the geometry of the flake. The complete formula for the energy of a polygonal flake is (see SI A.4 for derivation details):

$$U = \frac{4S'V_0(z)}{a_s^2} + \frac{2V_1(z)}{a_s^2}\left[\cos\left(\frac{4\pi}{\sqrt{3}a}T_y\right)\Phi_1\left(R,N,\beta,\theta,\frac{\theta}{2}\right) + \sin\left(\frac{4\pi}{\sqrt{3}a}T_y\right)\Phi_2\left(R,N,\beta,\theta,\frac{\theta}{2}\right)\right.$$
$$+ \cos\left(\frac{4\pi}{\sqrt{3}a}\left(\frac{\sqrt{3}}{2}T_x + \frac{T_y}{2}\right)\right)\Phi_1\left(R,N,\beta,\theta,\frac{\theta}{2}-\frac{\pi}{3}\right) + \sin\left(\frac{4\pi}{\sqrt{3}a}\left(\frac{\sqrt{3}}{2}T_x + \frac{T_y}{2}\right)\right)\Phi_2\left(R,N,\beta,\theta,\frac{\theta}{2}-\frac{\pi}{3}\right)$$
$$\left. + \cos\left(\frac{4\pi}{\sqrt{3}a}\left(\frac{T_y}{2} - \frac{\sqrt{3}}{2}T_x\right)\right)\Phi_1\left(R,N,\beta,\theta,\frac{\theta}{2}+\frac{\pi}{3}\right) + \sin\left(\frac{4\pi}{\sqrt{3}a}\left(\frac{T_y}{2} - \frac{\sqrt{3}}{2}T_x\right)\right)\Phi_2\left(R,N,\beta,\theta,\frac{\theta}{2}+\frac{\pi}{3}\right)\right]. \quad (44)$$

The size and shape dependence are captured primarily by two boundary terms $\Phi_1$ and $\Phi_2$:

$$\Phi_1(R,N,\beta,\theta,\theta_{\mathrm{m}}) = \sum_{i=1}^{N} \frac{R^2\cos\frac{\pi}{N}}{|\boldsymbol{G}_{\mathrm{m}}|R\sin(\alpha_i - \theta_{\mathrm{m}})}\left\{\frac{1-\cos(|\boldsymbol{G}_{\mathrm{m}}|R\cos\gamma_1)}{|\boldsymbol{G}_{\mathrm{m}}|R\cos\gamma_1} - \frac{1-\cos(|\boldsymbol{G}_{\mathrm{m}}|R\cos\gamma_2)}{|\boldsymbol{G}_{\mathrm{m}}|R\cos\gamma_2}\right\}, \quad (45)$$

$$\Phi_2(R,N,\beta,\theta,\theta_{\mathrm{m}}) = \sum_{i=1}^{N} \frac{R^2\cos\frac{\pi}{N}}{|\boldsymbol{G}_{\mathrm{m}}|R\sin(\alpha_i - \theta_{\mathrm{m}})}\left\{\frac{\sin(|\boldsymbol{G}_{\mathrm{m}}|R\cos\gamma_1)}{|\boldsymbol{G}_{\mathrm{m}}|R\cos\gamma_1} - \frac{\sin(|\boldsymbol{G}_{\mathrm{m}}|R\cos\gamma_2)}{|\boldsymbol{G}_{\mathrm{m}}|R\cos\gamma_2}\right\}, \quad (46)$$

Here, $S'$ denotes the area of the flake and the angle $\alpha_i = \theta + \beta + (i-1)2\pi/N$ specifies the orientation of the $i$-th congruent triangle composing the polygon. $\theta_{\mathrm{m}}$ denotes the direction of moiré reciprocal lattice vector derived from Eq. (38), and $\theta_{\mathrm{m1}} = \theta/2$, $\theta_{\mathrm{m2}} = \theta/2 - \pi/3$, $\theta_{\mathrm{m3}} = \theta/2 + \pi/3$, $\gamma_1 = \alpha_i - \theta_{\mathrm{m}} - \pi/N$, $\gamma_2 = \alpha_i - \theta_{\mathrm{m}} + \pi/N$. The complete derivation of the polygonal boundary formulation is provided in SI A.4.

To validate our polygonal formulation, we compare the energy computed from the analytical expressions with direct atomistic summation using Eq. (11). Fig. 7(b) and (c) present a comparison result for two representative polygonal flakes (A: $R = 40$ nm, $N = 9$, $\beta = 12°$; B: $R = 80$ nm, $N = 16$, $\beta = 5°$), showing the energy density as a function of twist angle $\theta$ and the difference of sliding energy surface. In both cases, the analytical results (lines) exhibit excellent agreement with atomistic calculations (symbols), confirming the accuracy of our formulation.

The polygonal boundary terms provide the link between edge geometry and circular-domain energetics. The boundary terms $\Phi_1$ and $\Phi_2$ serve as polygonal generalizations of the circular Bessel function term $\Phi_c$ in Eq. (40), encoding how the polygon's shape modulates the total energy. In SI A.6, we prove that as the number of sides $N$ tends to infinity, $\Phi_1$ converges to the Bessel function term $\Phi_c$, while $\Phi_2$ vanishes. Thus, the general polygonal formulation naturally reduces to the circular case of Eq. (40).

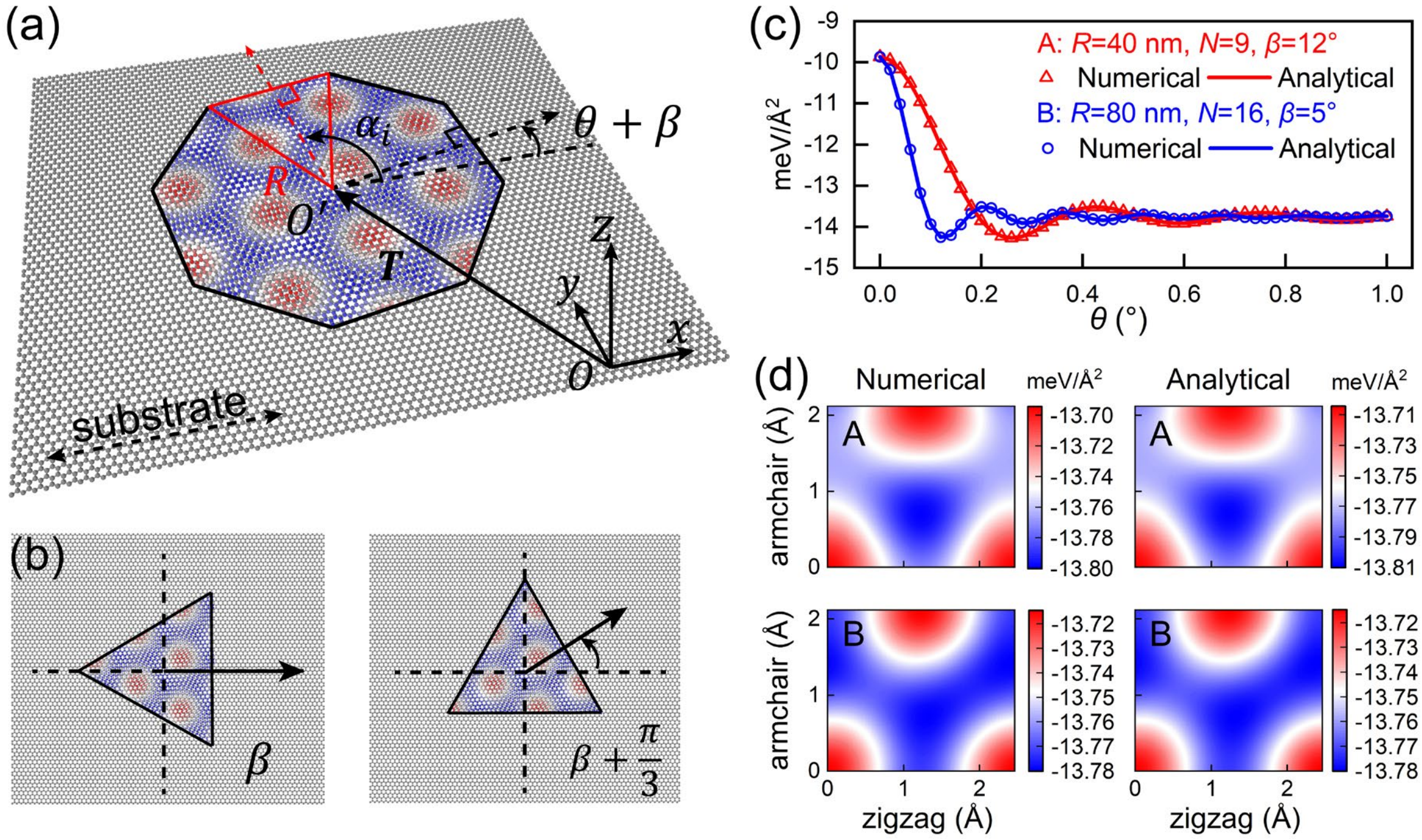

Fig. 7. Regular Polygonal Graphene Flake. (a) Key parameters are defined: the flake center $O'$, its position $\boldsymbol{T}$, the circumradius $R$, the polygonal boundary $\partial A'$, and the location angle of the $i$th isosceles triangle. The inset in the lower left corner shows two triangular flakes with different edge orientations under the same twist angle. (c) The energy density versus twist angle and (d) sliding energy surface with $\theta = 0.5°$ for polygonal graphene flakes. Results for two polygonal types (A: $R = 40$ nm, $N = 9$, $\beta = 12°$ and B: $R = 80$ nm, $N = 16$, $\beta = 5°$), calculated using the full analytical formula and direct summation based on Eq. (11) at $z = 3.35$ Å are presented.

In previous works [14,15], Yan et al. theoretically analyzed the relationship between the friction force and the boundary shape. Their conclusions are that, for a circular flake, the friction force scales with area to the 0.25 power during sliding, whereas for a polygonal flake, it exhibits a biperiodic oscillation, corresponding to a zero-scaling exponent. However, as the number of sides of a regular polygon increases, its shape gradually approaches that of a circle. The common boundary representation allows the change in scaling behavior to be examined as the polygon approaches its circular limit.

To establish a quantitative connection between boundary conditions, as a closed-form solution is unattainable, we numerically evaluate the intractable summation in Eq. (45) and compare it with the circular boundary's Bessel term in Eq. (40). We systematically investigate the effects of four key parameters: the number of polygon sides $N$, the interlayer twist angle $\theta$, the edge orientation $\beta$, and the flake size $R$ (defined by the circumradius). The numerical comparisons, summarized in Fig. 8, reveal several important trends:

(i) For a fixed flake size R, increasing the number of sides $N$ leads to a distinct transition in the difference between $\Phi_1$ and $\Phi_c$. Below a critical side number $N_{\mathrm{cr}}$, the difference exhibits strong oscillations; beyond $N_{\mathrm{cr}}$, it converges smoothly.

(ii) For a fixed sides number $N$ and generic edge orientations (i.e., those not aligned with the moiré periodicity), the polygonal result approximates the circular case only for small flake sizes. As $R$ increases, $\Phi_1$ oscillates within a bounded range without exhibiting a monotonic scaling law;

(iii) Although the scaling behaviors with size differ between circular and polygonal boundaries, their results remain very close to each other within a certain range of sizes, which is shown in Fig. 8, at $N = 25$, $\theta = 0.6°$, the difference between the two types of boundaries is clearly visible in the range of 1.5-2.0 nm, whereas in the range of 0-0.5 nm the results of the two curves are very close. The width of this range grows with side number $N$ increasing or with twist angle $\theta$ decreasing.

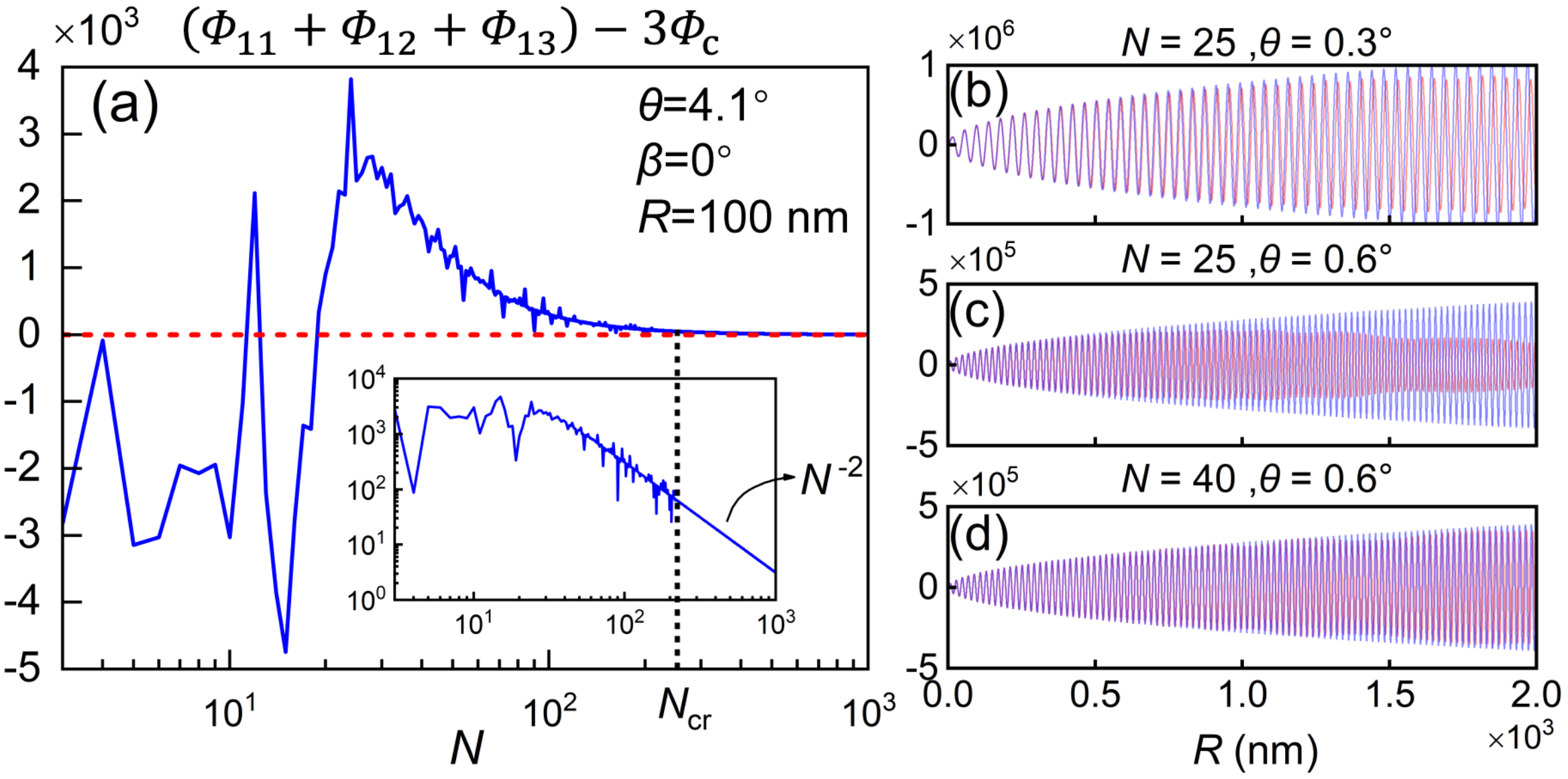

Fig. 8. Comparison of the boundary term calculations for regular polygons and circles. (a) Convergence trend with the number of sides $N_{\mathrm{cr}}$ (inset: log-log scale). A quadratic (2nd-order) power-law convergence emerges for $N > N_{\mathrm{cr}}$. (b-d) The convergence regime, where $\sum_i^3 \Phi_{1i}$ (red curve) $\approx 3\Phi_{\mathrm{c}}$ (blue curve), is enlarged by either increasing $N$ or reducing the twist angle $\theta$.

These trends identify two regimes in the approach to the circular limit. The behavior in (i) arises from two competing effects. First, as $N$ increases, the evolving polygon boundary sequentially incorporates new atoms, resulting in substantial numerical oscillations. Second, once the boundary aligns closely with the circumcircle, the physical atomic distributions inside the boundary become nearly identical. The remaining discrepancy stems from the non-physical energy contribution (i.e., an artifact of the first-order Fourier approximation), arising from integrating over the atom-free region between the polygon and the circle. This artifact decays gradually with increasing side count, manifesting as smooth convergence, with the crossover point $N_{\mathrm{cr}}$ marking the transition. Observations of (ii) and (iii) reflect the interplay of the following geometric quantities: the flake size $R$, the moiré period $a_{\mathrm{m}}$, and the polygon discretization. When $a_{\mathrm{m}} \gg R$ (small angles or tiny flakes), the energy density is nearly uniform, and the polygonal and circular boundaries appear similar. When $a_{\mathrm{m}} \sim R$, the polygon's discrete symmetry induces deviation, remedied by increasing side count to better approximate a circle. However, when $a_{\mathrm{m}} \ll R$, even a polygon with a large number of sides can still clearly resolve the variations of the moiré lattice along its boundary, so the scaling behavior of the polygon differs significantly from that of the circular boundary.

A dimensionless parameter combines the polygon's departure from its circumcircle with the moiré wave number:

$$\xi = |\boldsymbol{G}_{\mathrm{m}}| R \sqrt{1 - \cos\frac{\pi}{N}} \tag{47}$$

where $|\boldsymbol{G}_{\mathrm{m}}| = \frac{4\pi}{\sqrt{3}a_{\mathrm{m}}}$ is the magnitude of the moiré reciprocal lattice vector. This parameter combines the moiré wavelength ($2\pi/|\boldsymbol{G}_{\mathrm{m}}|$), flake size ($R$), and polygon discretization ($N$) into a single descriptor. By evaluating $\xi$ at the critical side number $N_{\mathrm{cr}}$ across a range of flake sizes and twist angles, we obtain the data shown in Fig. 9 (left panel). Across the conditions examined, $\xi$ converges to a value of approximately 2.17 across widely varying conditions. In the small-angle, small-size limit ($a_{\mathrm{m}} \gg R$), where polygon and circle are expected to converge, $\xi$ takes smaller values, consistent with our physical picture.

Adopting $\xi = 2$ as a practical threshold, Eq. (47) provides a predictive criterion for the polygon-to-circle transition. Given any two of the three parameters $(N, R, \theta)$, one can determine the critical value of the third at which the transition occurs. In general, when $\xi < 2$, the polygonal and circular boundaries follow the same size scaling law, whereas when $\xi > 2$, each follows its own distinct boundary-dependent behavior. While this criterion introduces some error near the origin (small angles or small sizes) where the calculated $\xi$ values are intrinsically small, it separates the regimes across most of the parameter range examined. As demonstrated in Fig. 9, the critical size $R_{\mathrm{cr}}$ predicted by the criterion for given side numbers and twist angles successfully separates the regimes: for $R < R_{\mathrm{cr}}$, the curves from both boundaries are in close agreement. The consistent convergence of $\xi$ to $\sim$2.17 across a wide parameter range provides a common numerical reference for the crossover. Elucidating its physical and mathematical roots presents a compelling direction for future investigation. Since $\Phi_2$ vanishes in the polygon-to-circle transition, the analysis above focused primarily on $\Phi_1$.

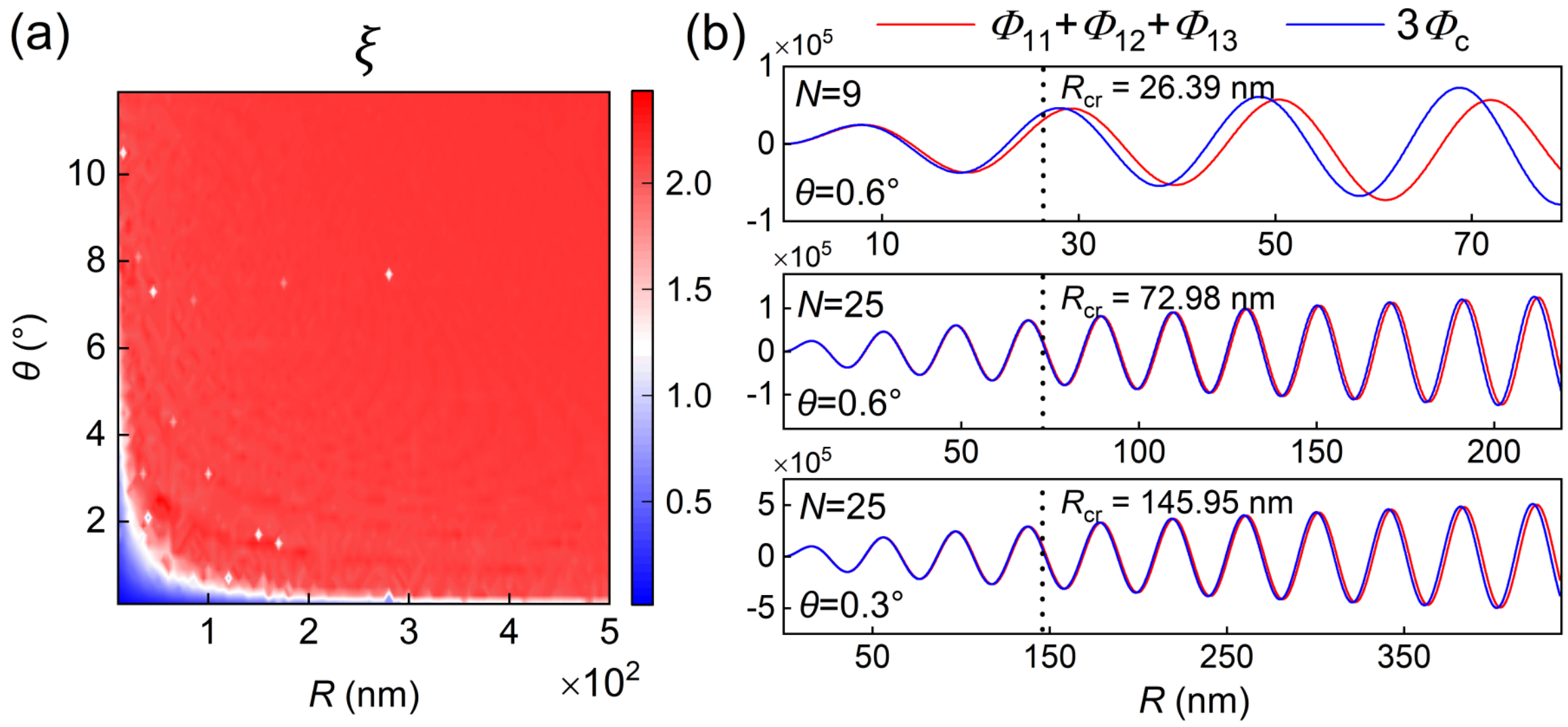

Fig. 9. Determination and validation of the transition parameter $\xi$. (Left) The values of $\xi$ are determined from the critical side number $N_{\mathrm{cr}}$ measured across different flake sizes and twist angles. (Right) Using $\xi = 2$ as a criterion, Eq. (47) predicts the critical size $R_{\mathrm{cr}}$ for the polygon-to-circle transition for a given side number and twist angle. For flakes smaller than $R_{\mathrm{cr}}$ ($\xi < 2$), the polygonal and circular boundaries yield nearly identical energy curves.

### *2.3.3 Limiting Cases and Asymptotic Behavior*

Both the circular and polygonal formulations converge in the limits of zeros twist angle ($\theta \to 0$) and infinite flake size ($R \to \infty$). However, the three boundary terms, $\Phi_{\mathrm{c}}$, $\Phi_1$ and $\Phi_2$ are functions of $\theta$ and $R$, and do not possess a unique double limit, the result depends critically on the order in which the limits $R \to \infty$ and $\theta \to 0$ are taken.

In the limit $\theta \to 0$, the moiré pattern vanishes. The total energy, which depends solely on the stacking registry (AA or AB), scales linearly with the flake area, which means $\Delta U \propto A^1$,

$$\lim_{\theta\to 0} \Phi_{\mathrm{c}} = \pi R^2, \qquad \lim_{\theta\to 0} \Phi_1 = NR^2 \sin\frac{\pi}{N}\cos\frac{\pi}{N}, \qquad \lim_{\theta\to 0} \Phi_2 = 0 \tag{48}$$

In the opposite limit $R \to \infty$, the total energy diverges due to the constant Fourier component. We therefore examine the areal energy density instead, in which the Fourier constant remains and boundary terms vanish, independent of the sliding position $\boldsymbol{T}$:

$$\lim_{R\to\infty} \frac{\Phi_{\mathrm{c}}}{R^2} = 0, \qquad \lim_{R\to\infty} \frac{\Phi_1}{R^2} = \lim_{R\to\infty} \frac{\Phi_2}{R^2} = 0, \qquad w = \frac{4V_0(z)}{a_s^2} \tag{49}$$

The cancellation of lateral forces in incommensurate interfaces is often conceptualized using the notions of moiré tiles (MT) and moiré rims (MR) [14,16,29,51]. Here, a MT represents a complete moiré period, whereas MR corresponds to an incomplete moiré region at the boundary. In calculations of friction force or energy barriers, the contributions from the MT often cancel out, leaving the incomplete MR as the dominant factor. The area integral treats MT and MR contributions within the same energy density. The integral over the MT regions vanishes automatically in the area integral of Eq. (26). Consequently, the remaining boundary terms, namely, $\Phi_{\mathrm{c}}$, $\Phi_1$ and $\Phi_2$, characterize the interlayer energy contribution governed primarily by the MR. In the infinite-flake limit, cancellation yields a constant energy density, implying sliding-invariant energy and thus superlubricity, as all MR vanish.

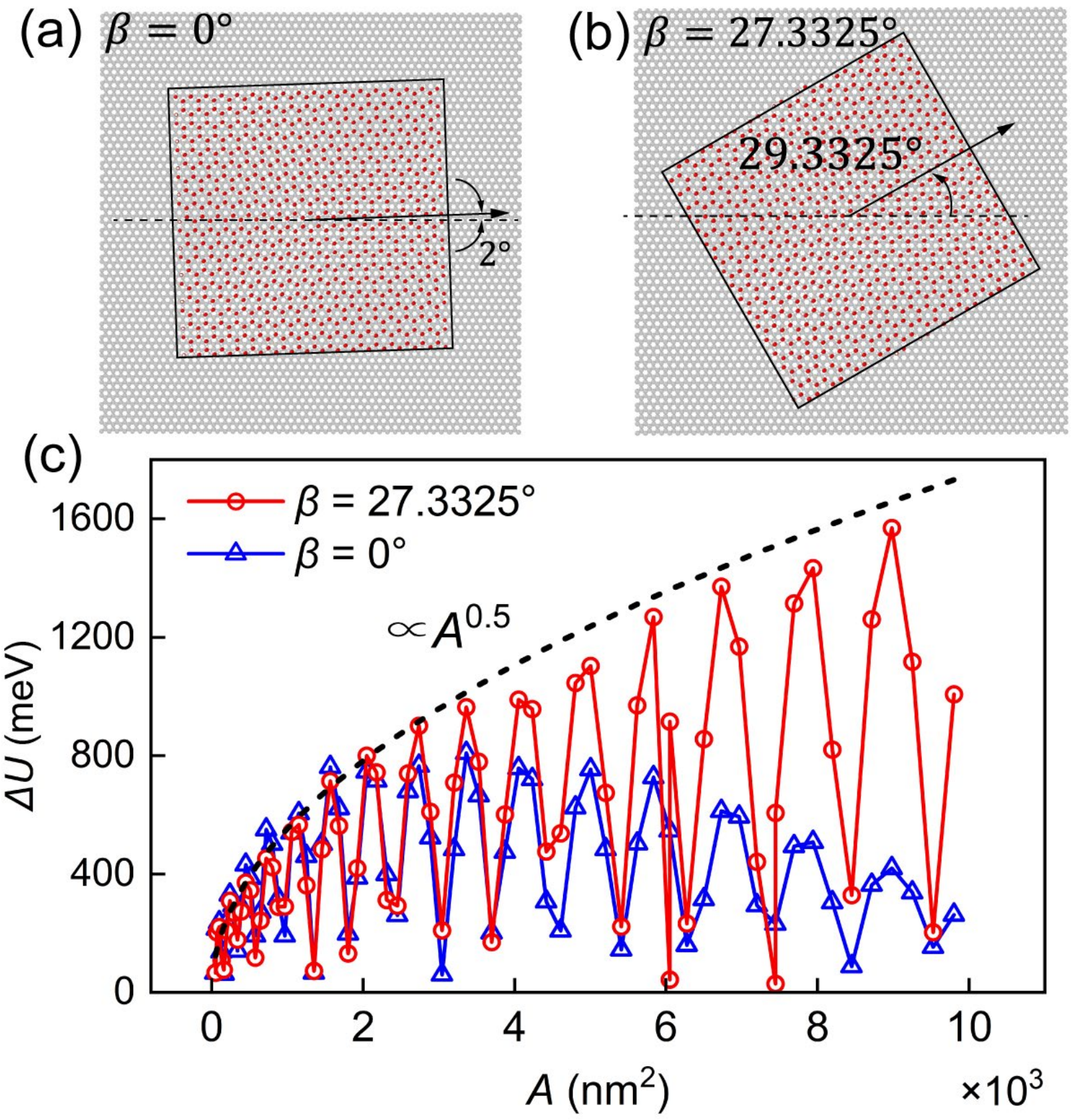

Fig. 10. Barrier amplitude of square 2D ice flake at $\theta = 2°$. (Top) atomic configuration of the square flake ($N = 4$) at two orientations of its edges, $\beta = 0°$ and $\beta = 27.3325°$; the latter fulfils the alignment condition $\alpha_i \to \theta_\mathrm{m}$ at this angle. (Bottom) sliding barrier amplitude $\Delta U$ against $A$ for the two orientations, on linear scales. For the generic orientation the amplitude oscillates with size and shows no monotonic scaling law, whereas for the aligned orientation its envelope follows $\Delta U \propto A^{0.5}$.
Square ice flakes test whether the generic and edge-aligned boundary regimes also appear in a molecular lattice, using the same geometric parametrization as for polygonal tBLG flakes. A square ($N = 4$) ice flake with orientation $\beta = 0°$ at twist angle $\theta = 2°$ is considered first. Its barrier amplitude oscillates with size and follows no monotonic scaling law (blue curve in Fig. 10), consistent with the behavior of polygonal graphene flakes described above. We then select the boundary orientation that satisfies the alignment condition $\alpha_i \to \theta_\mathrm{m}$ at the same twist angle; for $\theta = 2°$ gives $\beta = 27.3325°$. The barrier amplitude of the aligned square now follows the predicted power-law envelope, $\Delta U \propto A^{0.5}$ (red curve in Fig. 10), as predicted by the theory above. These two results reproduce, for a two-dimensional ice flake, the limiting behaviors derived in this section: no scaling law for a generic boundary orientation, and $A^{0.5}$ once the edges are aligned with the moiré lattice.

For polygonal boundaries, MR exhibits a distinct contribution mechanism, characterized by the limit $\alpha_i \to \theta_\mathrm{m}$, here we evaluate the limit of the general term within the summation for $\Phi_1$ and $\Phi_2$:

$$\begin{aligned}
\Phi_{1i} &= \lim_{\alpha_i \to \theta_{\mathrm{m}}} \Phi_1 = R\frac{2\sin\frac{\pi}{N}}{|\boldsymbol{G}_{\mathrm{m}}|}\sin\left(|\boldsymbol{G}_{\mathrm{m}}|R\cos\frac{\pi}{N}\right) + 2\sin\frac{\pi}{N}\frac{\cos\left(|\boldsymbol{G}_{\mathrm{m}}|R\cos\frac{\pi}{N}\right)-1}{|\boldsymbol{G}_{\mathrm{m}}|^2\cos\frac{\pi}{N}} \\
\Phi_{2i} &= \lim_{\alpha_i \to \theta_{\mathrm{m}}} \Phi_2 = R\frac{2\sin\frac{\pi}{N}}{|\boldsymbol{G}_{\mathrm{m}}|}\cos\left(|\boldsymbol{G}_{\mathrm{m}}|R\cos\frac{\pi}{N}\right) - 2\sin\frac{\pi}{N}\frac{\sin\left(|\boldsymbol{G}_{\mathrm{m}}|R\cos\frac{\pi}{N}\right)}{|\boldsymbol{G}_{\mathrm{m}}|^2\cos\frac{\pi}{N}}
\end{aligned}. \tag{50}$$

When summing the contributions of each edge of the polygon in Eq. (45) and Eq. (46), if none of the terms satisfies the limit condition $\alpha_i \to \theta_{\mathrm{m}}$, the resulting sum reduces to an oscillatory term in $R$, exhibiting no size scaling. In contrast, if one edge of the polygon satisfies this limit condition, that edge will provide a contribution linearly dependent on $R$, namely a scaling law of $A^{0.5}$, which is consistent with the molecular dynamics (MD) results [15,51]. The limiting terms in Eq. (50) which carries the factor $\sin\frac{\pi}{N}$ will vanish when $N \to \infty$, so the contribution linear in $R$, namely the $A^{0.5}$ branch disappears, and the summation for the rest $\Phi_{1i}$ reduces to $\Phi_{\mathrm{c}}$, resulting the scaling law of $A^{0.25}$, as observed in MD simulations [14,51].

The limit $\alpha_i \to \theta_{\mathrm{m}}$ is generally not accessible except under specific alignment: when a polygon edge is parallel to the moiré lattice direction[15]. Under this aligned condition, flake yields the $A^{0.5}$ scaling law until a larger side number induces a crossover to the $A^{0.25}$ scaling of a circular boundary.

## 2.4 Rotational Energetics of Finite Flakes

The angular derivative of the finite-domain energy connects boundary geometry to the rotational torque. Taking the negative derivative of the energy expressions Eq. (40) and Eq. (44) with respect to the twist angle yields the torque exerted by the substrate on the graphene flake. For a circular flake, the derivative has the following analytical form; polygonal torques are evaluated numerically below:

$$\begin{aligned}
M &= -\frac{\partial U}{\partial \theta} \\
&= \frac{4\pi R^2}{a_s a_s'} V_1(z)\left[\cos\left(\frac{4\pi}{\sqrt{3}a}T_y\right) + 2\cos\left(\frac{2\pi}{a}T_x\right)\cos\left(\frac{2\pi}{\sqrt{3}a}T_y\right)\right] J_2\left(\frac{4\pi R}{\sqrt{3}a_{\mathrm{m}}}\right)\frac{\cot\frac{\theta}{2}}{2},
\end{aligned} \tag{51}$$

here $J_2$ denotes the second-order Bessel function of the first kind. Using the sliding term in the formula, the center of the graphene flake is positioned at local AA ($T_x = T_y = 0$) and AB ($T_x = a/2,\ T_y = a/(2\sqrt{3})$) stacking points, and the torque exerted by the substrate is calculated under applied rotation. The torque is normalized by the area. For the polygonal flake, the torque is obtained from the variation of its energy with respect to the twist angle using a 4$^{\text{th}}$-order central difference scheme to compute the first derivative with respect to the radian angle.

Fig. 11 shows the torque on graphene flakes under rotation for two local stacking configurations. Both the circular flake radius and the polygonal flake circumradius are 15 nm, with the polygon orientation $\beta = 0°$. At zero twist, AA stacking is the highest-energy state; any finite twist induces a torque in the same direction as the rotation, driving the flake away from AA. Hence, the left panel shows positive torque at small angles. For AB stacking, which is an energy minimum, the torque opposes the rotation, as seen in the right panel.

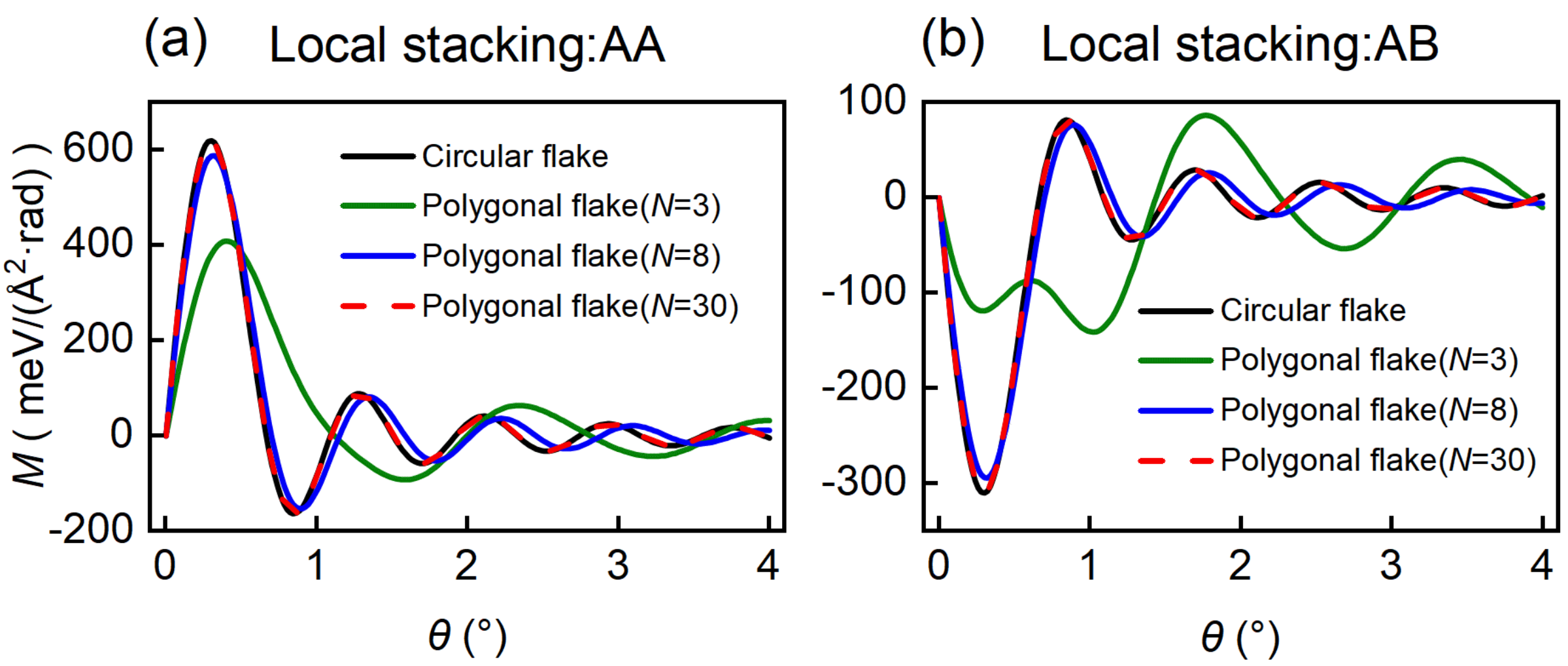

Fig. 11. Torque calculations for circular and polygonal graphene flakes. The radius of the circular flake and the circumradius of the polygonal flake are both set to 15 nm, with the orientation angle of the polygon $\beta = 0°$. The left panel corresponds to the local stacking configuration at AA stacking, while the right panel corresponds to AB stacking.

Equation (51) relates flake size to the spacing of rotational stationary states and the maximum torque. First, the zeros of $J_2(x)$ locate the extrema of $U(\theta)$ and therefore translate into a discrete spectrum of quasi-stable twist angles. In the small-angle limit, $x = \frac{4\pi R}{\sqrt{3}a_m} = \frac{4\pi R\theta}{\sqrt{3}a}$, so that

$$\theta_k = \frac{j_{2,k}\sqrt{3}a}{4\pi R} \propto \frac{1}{R}, \tag{52}$$

with $j_{2,k}$ the zeros of $J_2$. The spacing between successive zero points is $\Delta\theta_k = \theta_{k+1} - \theta_k \propto 1/R$, since the stationary points alternate between maxima and minima, the stable states occupy every second one, and their spacing is likewise $\propto 1/R$. A larger flake therefore accommodates a denser spectrum of stationary points (i.e., a denser set of stable states) within any fixed interval of twist angles. This size dependence of the density of stable states is consistent with the MD study of [52] on rotational-stability, in which finite twisted flakes untwist through a sequence of discrete, locally stable rotation states whose number scales with the flake size. The stable angles of successive orders decrease accordingly with increasing flake size, consistent with the analytical scaling law reported for the rotational energy landscape of twisted bilayers [53]. Second, in the small-angle limit, Eq. (51) reduces to $M \propto R^3 J_2(x)/x$, with $x = \frac{4\pi R\theta}{\sqrt{3}a}$, its angular factor $J_2(x)/x$ has the same form as the rotational friction torque $\tau \propto J_2(z)/z$ of [54] (with $z = x$ in this limit). The first maximum of $J_2(x)/x$ occurs at $x = x^* \approx 2.268$. As $R$ increases, reaching this maximum therefore requires the critical twist angle to scale as $\theta_{\mathrm{c,max}} = \frac{x^*\sqrt{3}a}{4\pi R} \propto R^{-1}$, so as to keep the Bessel argument fixed at $x = x^*$; it is precisely this $R^{-1}$ scaling that makes the normalized torque linear in $R$. Indeed, with $x = x^*$ held fixed (where $J_2(x^*)/x^*$ is a constant), the total maximum torque is $M_{\max} \propto R^3$, and dividing by the atom number $N \propto R^2$ gives the per-atom torque $\frac{M_{\max}}{N} \propto R$, in agreement with the $\tau_{\max} \propto R$ and $\theta_{\mathrm{c,max}} \propto R^{-1}$ scalings reported by [54].

# 3 Applications and discussion

Energy gradients and curvatures translate the finite-domain energy, Eq. (40), into shear stress and local stiffness. Because the same boundary terms enter both quantities, the energy formulation also determines how contact shape, size and twist angle influence the interlayer mechanical response.

## 3.1 Size Scaling of Shear Stress and Shear Modulus

The interlayer shear response follows from derivatives of the finite-flake energy. To recover the correct AB-stacking modulus at $\theta \to 0$, the coordinate of the upper flake is translated so that its hexagonal lattice center coincides with a substrate atom, creating a local AB configuration, namely, $T_x = \frac{a}{2}$, $T_y = \frac{a}{2\sqrt{3}}$. The interlayer shear modulus $G$ is defined by $\tau = G\gamma$ , with the shear strain $\gamma = T_x/z$, where $T_x$ is the sliding displacement along x direction (zigzag) and $z$ is the interlayer distance [55,56]. Meanwhile, the shear stress $\tau$ is obtained by dividing the force acting on the entire graphene flake by its area. The force on the flake itself can be derived by calculating the negative gradient of energy (see SI A.5 for details).

Because the energy is periodic in $\boldsymbol{T}$ and the substrate is taken as rigid, the modulus has to be evaluated as the curvature at the equilibrium registry rather than at an arbitrary translation. Writing the sliding energy as $U = \frac{4\pi R^2 V_0}{a_s^2} + \frac{2V_1}{a_s^2} P_T(\boldsymbol{T})\Phi_{\mathrm{c}}$, the internal restoring stress is $\tau_{\mathrm{int}} = -A^{-1}\mathrm{d}U/\mathrm{d}T$, and the positive tangent modulus at equilibrium is $G = zA^{-1}\mathrm{d}^2U/\mathrm{d}T^2 = -z\mathrm{d}\tau_{\mathrm{int}}/\mathrm{d}T$; evaluating the curvature at a fixed but non-equilibrium translation would instead return a registry-constrained stiffness

For circular flake

$$\tau = \frac{4V_1(z)}{a_s^2}\frac{4\pi}{\sqrt{3}a}\frac{J_1\left(\frac{4\pi R}{\sqrt{3}a_{\mathrm{m}}}\right)}{\frac{4\pi R}{\sqrt{3}a_{\mathrm{m}}}} f_{\mathrm{zig}}(T_x, T_y), \tag{53}$$

Where the factor $\frac{4\pi}{\sqrt{3}a}\boldsymbol{f}(T_x, T_y) = \frac{4\pi}{\sqrt{3}a}(f_{\mathrm{zig}}, f_{\mathrm{arm}})$ is obtained by differentiating the translation term of Eq. (40). Writing $P_T$ in Cartesian form, $P_T = \cos\left(\frac{4\pi T_y}{\sqrt{3}a}\right) + 2\cos\left(\frac{2\pi T_x}{a}\right)\cos\left(\frac{2\pi T_y}{\sqrt{3}a}\right)$, its negative gradient gives

$$f_{\text{zig}}(T_x,T_y)=\sqrt{3}\sin\left(\frac{2\pi}{a}T_x\right)\cos\left(\frac{2\pi}{\sqrt{3}a}T_y\right)$$
$$f_{\text{arm}}(T_x,T_y)=\sin\left(\frac{4\pi}{\sqrt{3}a}T_y\right)+\cos\left(\frac{2\pi}{a}T_x\right)\sin\left(\frac{2\pi}{\sqrt{3}a}T_y\right) \tag{54}$$

Evaluating the tangent response at the equilibrium registry gives the shear modulus:

$$G=\left.\frac{d\tau}{d\gamma}\right|_{AB}=z\left.\frac{d\tau}{dT_x}\right|_{AB}=\sqrt{3}z\frac{\pi}{a}\frac{4\pi}{\sqrt{3}a}\frac{4V_1(z)}{a_s^2}\frac{J_1\left(\frac{4\pi R}{\sqrt{3}a_{\text{m}}}\right)}{\frac{4\pi R}{\sqrt{3}a_{\text{m}}}}. \tag{55}$$

The sign of the expression is governed by $\frac{J_1(x)}{x}$ and indicates whether the local AB registry is a stable minimum or a local maximum of the sliding landscape; where it becomes negative, AB is no longer the equilibrium registry, so that $|G|$ measures the local stiffness scale rather than the modulus of a relaxed configuration. Its size dependence, however, is set by the envelope of $\frac{J_1(x)}{x}$: because $\frac{J_1(x)}{x}$ oscillates about zero as the size varies, only the envelope – the upper bound of the oscillating stiffness – is physically meaningful, and it is the magnitude $|G|$ that is analyzed and plotted below. The results presented here are calculated for sliding along the zigzag direction. In the limit $\theta\to 0$, the Bessel function term converges to $0.5$ and we obtain the shear modulus for bilayer graphene in AB stacking, with a value of $4.54$GPa, which is close to previous work, around $4.5$GPa [57-59].

The size dependence of Eq. (55) is contained in the Bessel factor. Using the asymptotic envelope

$$\left|\frac{J_1(x)}{x}\right|\sim\sqrt{\frac{2}{\pi}}x^{-\frac{3}{2}}\propto R^{-\frac{3}{2}}\theta^{-\frac{3}{2}},\qquad\left(x=\frac{4\pi R}{\sqrt{3}a_{\text{m}}}\gg 1\right), \tag{56}$$

Consequently, the interlayer shear stiffness of a finite twisted flake oscillates and its envelope decays to zero with increasing size as shown in Fig. 12. The incommensurate thermodynamic limit has vanishing shear stiffness, which is the stiffness formulation of structural superlubricity. This is consistent with MD calculations in which the twisted interlayer shear modulus drops from ~2 GPa (commensurate) to 5-40MPa (incommensurate) [60]. A further quantitative comparison uses the interlayer force constant $\alpha_{\parallel}=12.8\times 10^{18}\ \text{N/m}^3$ extracted from Raman C-mode measurements [61], which translates to $G=\alpha_{\parallel}d_0\approx 4.3$ GPa. Equivalently, using the linear chain model (LCM) [59,61], our force constant $\alpha_{\parallel}=\frac{G}{d_0}=13.6\times 10^{18}\ \text{N/m}^3$ predicts the Raman C-mode frequencies of bulk graphite $\omega_C^{\text{bulk}}=\frac{1}{\pi c}\sqrt{\frac{\alpha_{\parallel}}{\mu}}=44.8\ \text{cm}^{-1}$ and bilayer graphene $\omega_C^{2\text{L}}=\frac{\omega_C^{\text{bulk}}}{\sqrt{2}}=31.7\ \text{cm}^{-1}$, close to the experimental measured values of 43.5 and 31 $\text{cm}^{-1}$[61].

We can further investigate the interlayer shear strength using Eq. (53). Under the condition that the system remains locally AB-stacked and slides along the zigzag direction, setting $T_y = \frac{a}{2\sqrt{3}}$, the maximum stress is given by:

$$\tau_{\max} = \left| \frac{4V_1(z)}{a_s^2} \frac{4\pi}{\sqrt{3}a} \frac{J_1\left(\frac{4\pi R}{\sqrt{3}a_{\mathrm{m}}}\right)}{\frac{4\pi R}{\sqrt{3}a_{\mathrm{m}}}} \frac{\sqrt{3}}{2} \sin\left(\frac{2\pi}{a} T_x\right) \right|_{\max} \tag{57}$$

Substituting the numerical values and taking $\theta \to 0$ yields a maximum stress of approximately 0.54 GPa for AB stacking. A first-principles calculation gives 0.38 GPa in [62], while the experimental value is 62 MPa [63]. Under external loading, dislocations, strain solitons and elastic deformation develop at the interface [64,65], These structures provide additional pathways for shear [63,66,67], substantially reducing the cost for sliding. The present calculation characterizes the rigid, coherent sliding response associated with the interlayer registry barrier.

The curvature calculation used for Eq. (55) also gives the zigzag interlayer shear stiffness of a polygonal flake

$$G = \sqrt{3}z \frac{\pi}{a} \frac{4\pi}{\sqrt{3}a} \frac{4V_1(z)}{a_s a_s'} \frac{\frac{1}{2}(\Phi_{12} + \Phi_{13}) + \frac{\sqrt{3}}{2}(\Phi_{22} + \Phi_{23})}{2NR^2 \sin\frac{\pi}{N} \cos\frac{\pi}{N}}, \tag{58}$$

while $\Phi_{i2} = \Phi_i\left(R, N, \beta, \theta, \frac{\theta}{2} - \frac{\pi}{3}\right)$, $\Phi_{i3} = \Phi_i\left(R, N, \beta, \theta, \frac{\theta}{2} + \frac{\pi}{3}\right)$. Eq. (55) and Eq. (58) present the concrete expressions for the interlayer shear stiffness of circular and polygonal graphene flakes under zigzag-direction sliding. The two equations share an identical structure, formed by multiplying a specific coefficient with a boundary geometry term. Furthermore, the boundary term in Eq. (58) reduces to that of the circular case (Eq. (55)) as the number of sides tends to infinity. Fig. 12 maps the resulting stiffness as a function of flake size and twist angle. Since the stiffness is calculated by dividing by the flake's area, neither the circular nor polygonal flakes exhibit a scaling law with size. Instead, the stiffness oscillates and decays to zero as the size increases (except for $\theta = 0°$). A considerable number of points where the shear stiffness is zero can be observed in the figure. These zero-stiffness points occur when specific relationships between size and twist angle cause Eq. (40), (45) and (46) evaluate to zero. Within our model, these points correspond to cases of complete moiré complementation. In such a state, the interaction potential energy between the graphene flake and the substrate becomes a constant. Consequently, sliding the flake encounters no energy barrier. Panels (b, c) and (e, f) are line profiles extracted from panels (a) and (d), respectively. The line profiles resolve the dependence of shear stiffness on flake size and twist angle. Starting from the AB-stacked configuration, as the twist angle increases, the shear stiffness drops sharply, and this decrease becomes more pronounced for larger flakes. Similarly, for non-AB-stacked configurations, increasing the flake size also leads to a sharp reduction in shear stiffness. This size- and angle-dependent behavior is governed by the boundary terms. The boundary term measures the net contribution after cancellation across the flake. Perfect moiré registry is achieved when the flake size is as large as possible while the moiré period is as small as possible, i.e., $|\boldsymbol{G}_{\mathrm{m}}|R$ is large enough. In this regime, whether it is the sliding energy barrier, the static friction force, or the shear behavior discussed here, all are influenced by the boundary terms.

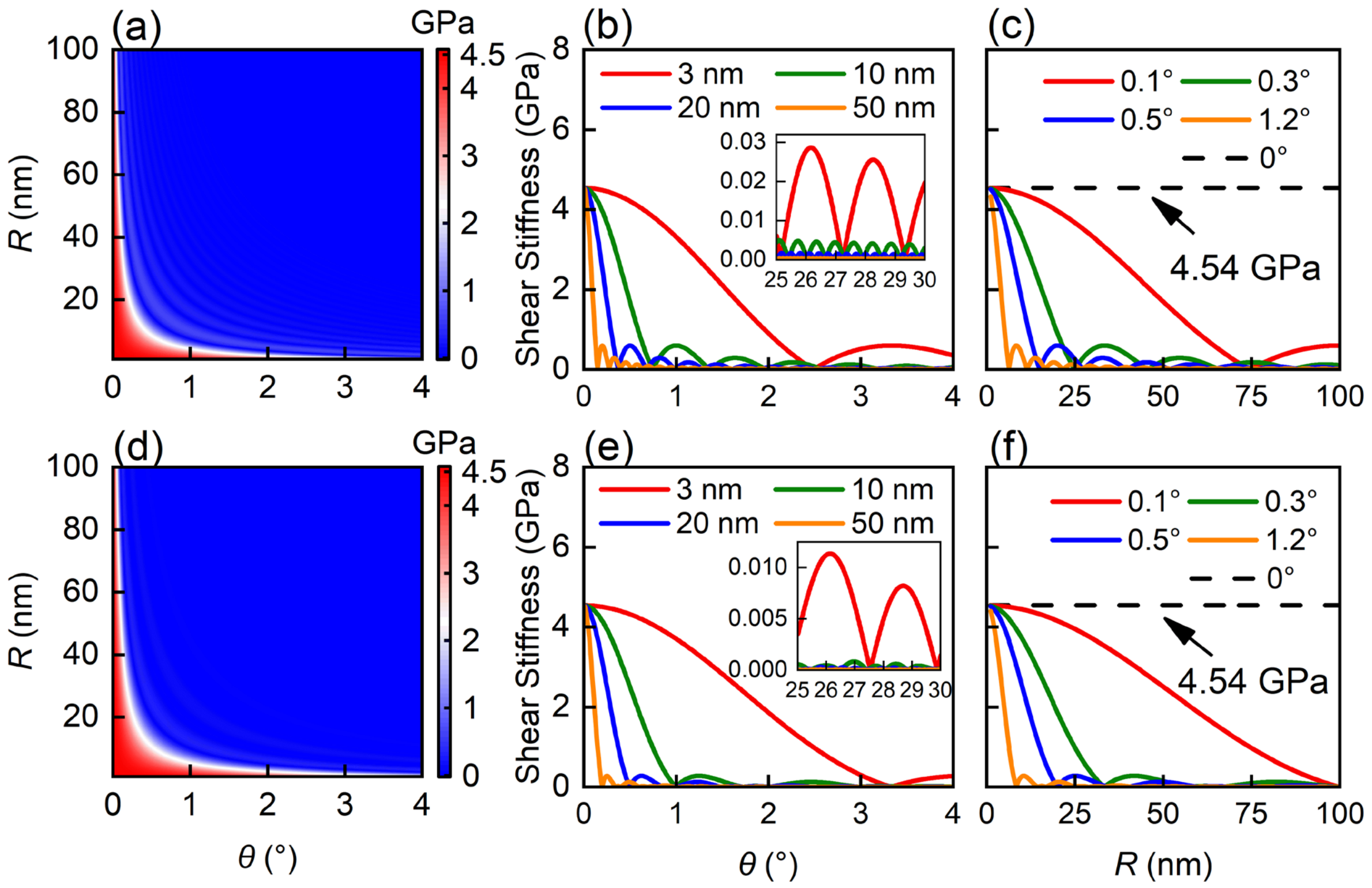

Fig. 12. The interlayer shear stiffness. (a-c) are results for circular flake and (d-f) for polygonal flake with $N = 4$ and $\beta = 0°$. (b), (c), (e), and (f) show line profiles extracted from the maps in (a) and (d), respectively. Their vertical axes represent the shear stiffness in GPa. In (c) and (f), the height of the black dashed line ($\theta = 0°$) corresponds to 4.54 GPa. The sign of the Eq. (55) which tells whether the AB registry is a minimum or a maximum, is not shown.

## 3.2 Mechanical behavior of Bulk Graphite under Compression

Eq. (33) and Eq. (34) also determine the interaction energy of an $N$-layer graphite stack. With one layer fixed at zero height, the remaining layers are placed at a uniform spacing $z$ with AB stacking between adjacent layers. To account for interactions between non-adjacent layers, the total potential energy density is expressed as:

$$w_{\text{total}} = \sum_{i=1}^{N}\sum_{j>i}^{N} w_{ij}\,, \qquad \text{where } w_{ij} = \begin{Bmatrix} w_{\text{AA}}(z_{ij}), & \text{parity}(i) = \text{parity}(j) \\ w_{\text{AB}}(z_{ij}), & \text{otherwise} \end{Bmatrix}, \tag{59}$$

Here $w_{\text{AA}} = w_1(\boldsymbol{T} = \boldsymbol{0})$, $w_{\text{AB}} = w_1\left(\boldsymbol{T} = \frac{1}{3}\boldsymbol{a}_1 + \frac{1}{3}\boldsymbol{a}_2\right)$ and $z_{ij} = (j-i)z > 0$. To further simplify this expression, we introduce the interlayer distance multiple $d = j - i$. Since the maximum value of $j$ is $N$, for a fixed $d$, there are $N - d$ possible value for $i$, then there are $N - d$ possible pairs for $(i, j)$. Since $d$ is fixed, the interactions for all $N - d$ pairs $(i, j)$ are identical, and the total contribution from these pairs is given by $(N-d)w_{\text{AA}}(dz)$ when $d$ is even and $(N-d)w_{\text{AB}}(dz)$ when $d$ is odd, namely

$$w_{\text{total}}(z; N) = \sum_{k=1}^{\left\lfloor\frac{N-1}{2}\right\rfloor} (N-2k)w_{\text{AA}}(2kz) + \sum_{k=1}^{\left\lfloor\frac{N}{2}\right\rfloor}\big(N-(2k-1)\big)w_{\text{AB}}\big((2k-1)z\big), \tag{60}$$

here the brackets $\lfloor * \rfloor$ represent the floor function. For clarity, we present the explicit expressions for the two types of stacking:

$$w_{\text{AA}}(z) = \frac{4V_0(z) + 6V_1(z)}{a_s^2}, \qquad w_{\text{AB}}(z) = \frac{4V_0(z) - 3V_1(z)}{a_s^2}. \tag{61}$$

The retained lattice information distinguishes AA and AB interactions. Averaging out the atomic structure would leave the zeroth-order Fourier component of the single-atom expression . In that case, $w_{\text{AA}}$ and $w_{\text{AB}}$ would become identical.

After obtaining the interaction potential energy density for an N-layer graphite block with AB stacking, we multiply it by the area $A$ to obtain the total energy. Taking the derivative of the total energy with respect to the interlayer distance $z$ yields the external force required to maintain the compressed (or stretched) configuration. Dividing this force by the area $A$ gives the uniform external stress applied to the graphite block:

$$(N-1)\sigma(z; N) = \frac{\mathrm{d}w_{\text{total}}(z; N)}{\mathrm{d}z}$$

$$= \sum_{k=1}^{\left\lfloor\frac{N-1}{2}\right\rfloor} 2k(N-2k)\left[\frac{\mathrm{d}w_{\text{AA}}(\eta)}{\mathrm{d}\eta}\right]_{\eta=2kz} + \sum_{k=1}^{\left\lfloor\frac{N}{2}\right\rfloor}(2k-1)\big(N-(2k-1)\big)\left[\frac{\mathrm{d}w_{\text{AB}}(\eta)}{\mathrm{d}\eta}\right]_{\eta=(2k-1)z}. \tag{62}$$

This expression explicitly gives the magnitude of the stress applied to the AB-stacked graphite block when the interlayer distance is $z$. Meanwhile, the equilibrium interlayer distance $z_{\text{eq}}$ for a given number of layers N can be obtained by numerically optimizing the stress to zero and results are shown in Fig. 13. In the equation, $N$-1 represents the actual displacement under external compression. For a graphite block consisting of $N$ layers, the total displacement is $N$-1 times the interlayer spacing change.

In accordance with the generalized Hooke's law and using the Voigt notation, the stress-strain relation of graphite is given by:

$$\begin{bmatrix} \sigma_{xx} \\ \sigma_{yy} \\ \sigma_{zz} \end{bmatrix} = \begin{bmatrix} C_{11} & C_{12} & C_{13} \\ C_{12} & C_{11} & C_{13} \\ C_{13} & C_{13} & C_{33} \end{bmatrix} \begin{bmatrix} \varepsilon_{xx} \\ \varepsilon_{yy} \\ \varepsilon_{zz} \end{bmatrix}, \tag{63}$$

since our theoretical model does not consider in-plane deformation, the in-plane strains can be set to zero, and we have $\sigma_{zz} = C_{33}\varepsilon_{zz}$, namely

$$\sigma(z; N) = C_{33}\left(\frac{z}{z_{\mathrm{eq}}} - 1\right). \tag{64}$$

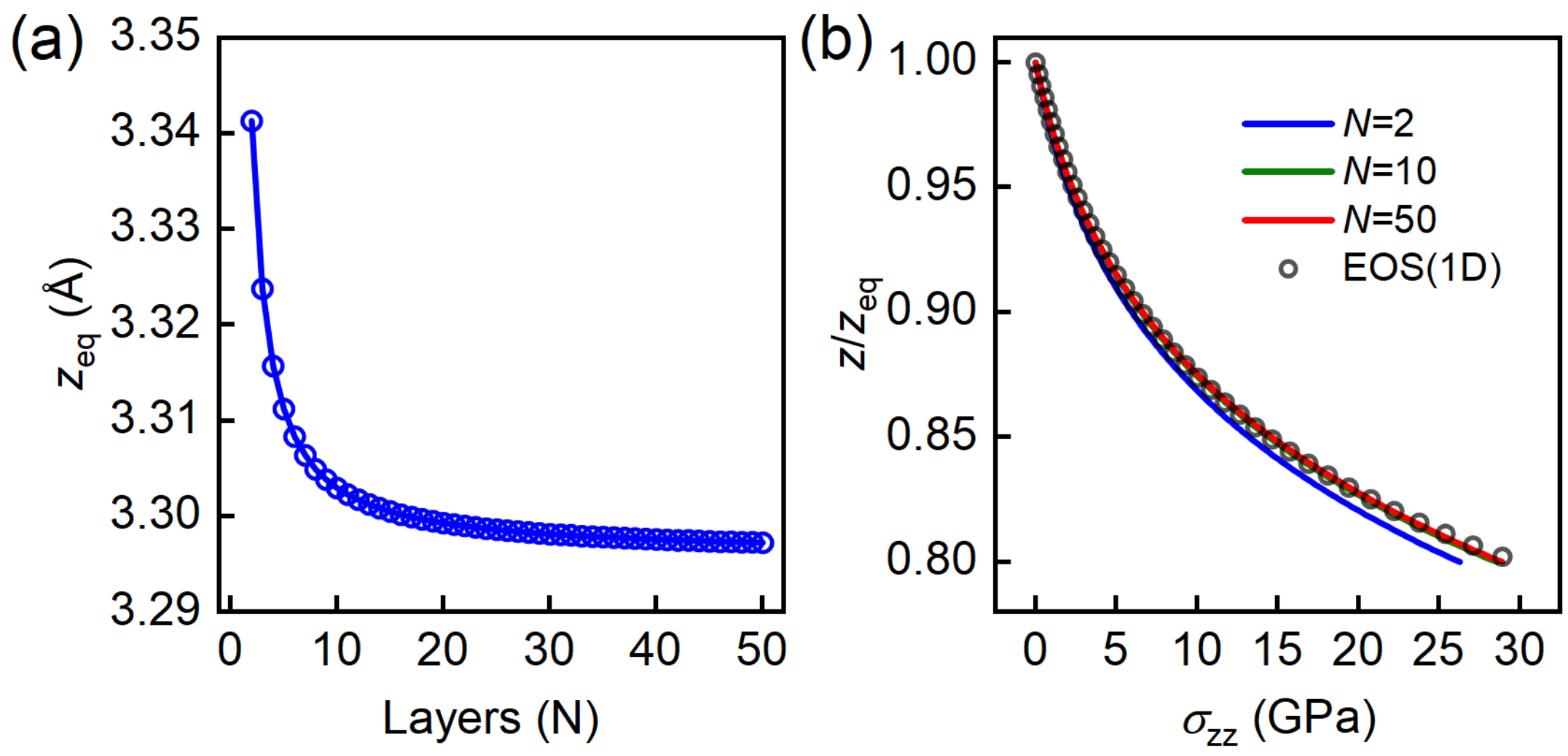

Fig. 13. Compression of graphite. Left panel: Equilibrium interlayer distances for graphite blocks with different numbers of layers, obtained by numerical optimization using theoretical model. Right panel: Compression behavior of graphite blocks with 2 layers (blue line), 10 layers (green line), and 50 layers (red line). The red points represent the fitting results obtained using the one-dimensional effective Murnaghan equation of state (EOS).

In Fig. 13 shows how layer number affects the equilibrium spacing and compression response of graphite. The interlayer spacing on the vertical axis is normalized by the corresponding equilibrium interlayer height $z_{\mathrm{eq}}$. The interlayer interactions in the graphite block can be treated as nonlinear springs. For a large number of layers, the contributions from non-adjacent layers gradually saturate as their separation exceeds the interaction range. Consequently, the compressive responses of the 10-layer and 50-layer blocks are very close to each other. According to the generalized Hooke's law, under the condition of zero in-plane strain, the proportional coefficient between $\sigma_{zz}$ and $\varepsilon_{zz}$ is precisely the elastic constant $C_{33}$. From the linear regime at the onset of compression in the right panel of Fig. 13, the $C_{33}$ value of bulk graphite of AB stacking is determined to be around 31.5 GPa, compared with the reported experimental value of 38.7±7 GPa [46] and slightly lower than the first-principles result 42 GPa [57]. These comparisons place the calculated response on the scale of the reported interlayer elastic constants. Considering the approximate nature of our model, the agreement is considered reasonable. In addition to the $C_{33}$ parameter, the bulk modulus can also be obtained from the compression curves shown in the Fig. 13. In general, the bulk modulus is determined by fitting the compression curve using the Murnaghan equation of state (EOS) [68,69]:

$$\frac{V(P)}{V} = \left[1 + \left(\frac{B_V'}{B_V^0}\right) P\right]^{-\frac{1}{B_V'}}, \tag{65}$$

from which both the bulk modulus $B_V^0$ and its derivative at zero pressure $B_V'$ can be extracted. However, because our energy-based theory does not account for in-plane deformation, the hydrostatic condition required in the above formula cannot be realized. We therefore adopt an equivalent one-dimensional version of the formula:

$$\frac{z(\sigma_{zz})}{z_0} = \left[1 + \left(\frac{B_V'}{B_V^0}\right) \sigma_{zz}\right]^{-\frac{1}{B_V'}}, \tag{66}$$

The fitting result of this formula for the 50-layer graphite is presented in Fig. 13. The equivalent bulk modulus along the z-direction is determined to be 34.63 GPa and a zero-pressure derivative of 10.23.

# 4 Conclusions

In summary, we have established a unified reciprocal-space energy-density framework connecting microscopic registry-dependent interactions to the contact-scale energetics and mechanics of van der Waals interfaces. By representing the substrate potential and overlayer atomic density in Fourier space, the formulation retains lattice symmetry while making the integration domain explicit, placing single-atom and molecular adsorption, infinite bilayers, and finite crystalline contacts on a common analytical footing.

For circular and regular polygonal contacts, the total interfacial energy separates into a bulk area contribution and boundary-dependent geometric terms. This unified representation connects the circular Bessel form to explicit polygonal boundary integrals, revealing how edge orientation determines the asymptotic scaling of the energy barrier. A dimensionless parameter combining moiré wavelength, flake size, and number of sides characterizes the polygon-to-circle crossover. Furthermore, calculations for two-dimensional ice based on water–graphene interaction parameters reproduce the corresponding boundary-scaling trends in a molecular lattice.

Energy derivatives connect the finite-domain expressions to rotational torque and interlayer shear stiffness. For circular flakes, the rotational energy exhibits a sequence of stationary angles with characteristic spacing inversely proportional to flake size. The interlayer shear stiffness agrees with Raman-derived interlayer force constants in the commensurate limit and captures the size- and twist-dependent reduction in stiffness toward incommensurate contacts. Extended to bulk graphite, the same microscopic interaction description yields the normal compression response and an out-of-plane elastic constant comparable to reported experimental values.

Within this common formulation, lattice structure, moiré periodicity, and contact geometry directly connect microscopic interactions to finite-contact energy landscapes and conservative mechanical responses. These relationships provide an analytical basis for predicting how lattice registry and geometric design shape the energetics and mechanics of van der Waals interfaces.

**CRediT authorship contribution statement**

Zhangke Lei: Software, Formal analysis, Methodology, Investigation, Validation, Writing - original draft. Weidong Yan: Formal analysis, Validation, Writing - review & editing. Zhicheng Feng: Software, Validation, Writing - review & editing. Wengen Ouyang: Conceptualization, Supervision, Methodology, Validation, Resources, Funding acquisition, Project administration, Writing - review & editing.

**Declaration of competing interest**

The authors declare that they have no known competing financial interests or personal relationships that could have appeared to influence the work reported in this paper.

**Acknowledgments**

We thank Prof. Ze Liu and Prof. Michael Urbakh for useful discussions on this topic. W.O. acknowledges the National Natural Science Foundation of China (12472099, U2441207), the Fundamental Research Funds for the Central Universities (No. 2042025kf0050), the National Key Research and Development Program of China (Nos. 2025YFF0513900 and 2025YFF0513901) and the start-up fund of Wuhan University. Computations were conducted at the Supercomputing Center of Wuhan University and the National Supercomputer TianHe-1(A) Center in Tianjin.